\documentclass[letterpaper]{article} 
\usepackage[]{aaai25}  
\usepackage{times}  
\usepackage{helvet}  
\usepackage{courier}  
\usepackage[hyphens]{url}  
\usepackage{graphicx} 

\usepackage{booktabs}
\usepackage{multirow}
\usepackage{amssymb} 
\usepackage{pifont}  
\usepackage{amsmath}
\usepackage{bm}
\usepackage{algorithm}
\usepackage{algpseudocode}

\usepackage{siunitx}  
\usepackage{xcolor}
\newcommand{\answerYes}[1]{\textcolor{blue}{#1}} 
 
\newcommand{\answerNA}[1]{\textcolor{gray}{#1}}

\usepackage{natbib}  
\usepackage{caption} 
\usepackage{newfloat}
\usepackage{listings}
\DeclareCaptionStyle{ruled}{labelfont=normalfont,labelsep=colon,strut=off} 
\floatstyle{ruled}
\newfloat{listing}{tb}{lst}{}
\floatname{listing}{Listing}
\title{Multi-source Multi-level Multi-token Ethereum Dataset and Benchmark Platform}

\title{Multi-source Multi-level Multi-token Ethereum Dataset and Benchmark Platform}  
\author { 
    Haoyuan Li\textsuperscript{\rm 1}, 
    Mengxiao Zhang\textsuperscript{\rm 1}, 
    Maoyuan Li\textsuperscript{\rm 1}, 
    Jianzheng Li\textsuperscript{\rm 1}, 
    Junyi Yang\textsuperscript{\rm 1}, 
    Shuangyan Deng\textsuperscript{\rm 1}, 
    Zijian Zhang\textsuperscript{\rm 2}, 
    Jiamou Liu\textsuperscript{\rm 1} 
} 

\affiliations { 
    \textsuperscript{\rm 1}School of Computer Science, The University of Auckland, Auckland, New Zealand\\ 
    \textsuperscript{\rm 2}School of Computer Science and Technology, Beijing Institute of Technology, Beijing, China\\ 
    hli962@aucklanduni.ac.nz, 
    jiamou.liu@auckland.ac.nz 
}

\usepackage{bibentry}

\begin{document}

\maketitle

\begin{abstract}
This paper introduces 3MEthTaskforce ({\bf \url{3meth.github.io}}), a multi-source, multi-level, and multi-token Ethereum dataset addressing the limitations of single-source datasets. Integrating over 300 million transaction records, 3,880 token profiles, global market indicators, and Reddit sentiment data from 2014–2024, it enables comprehensive studies on user behavior, market sentiment, and token performance. 3MEthTaskforce defines benchmarks for user behavior prediction and token price prediction tasks, using 6 dynamic graph neural networks and 19 time-series models to evaluate performance. Its multimodal design supports risk analysis and market fluctuation modeling, providing a valuable resource for advancing blockchain analytics and decentralized finance research.
\end{abstract}

%

\section{Introduction}

The cryptocurrency market has witnessed significant growth over the past decade, driven by the rapid development of Ethereum and the widespread adoption of the ERC-20 token standard. As of 2024, the total market capitalization of cryptocurrencies is approximately 1.5 trillion USD, with around 300 million individuals globally holding cryptocurrencies and approximately 10,025 active tokens in circulation\footnote{\url{https://etherscan.io/chart/marketcap}}\footnote{\url{https://www.demandsage.com/blockchain-statistics/}}. Ethereum alone processes an average of 1.2 million transactions per day\footnote{\url{https://ycharts.com/indicators/ethereum_transactions_per_day}}, highlighting the immense scale and complexity of the cryptocurrency ecosystem.

However, this market is characterized by high volatility, influenced by factors such as token prices, global market indicators, user behavior, and market sentiment. The UST de-pegging event in May 2022 \cite{briola2023anatomy,uhlig2022luna} triggered widespread panic and led to the collapse of LUNA (now LUNC), underscoring the profound impact of rapid sentiment shifts on related assets. Consequently, researchers have increasingly focused on leveraging machine learning to analyze user behavior and predict cryptocurrency prices \cite{khedr2021cryptocurrency,chen2020bitcoin,akila2023cryptocurrency,chen2023analysis}.

Despite growing interest in machine learning for cryptocurrency analysis, existing datasets are often fragmented, focusing on narrow aspects of the market. For instance, repositories such as Transaction Graph Dataset for the Ethereum Blockchain V2 (TGDEB) \cite{ozturan2021transaction} and Chartalist \cite{shamsi2022chartalist} provide transaction-level data, while platforms like Kaggle and Huggingface\footnote{\url{https://huggingface.co/datasets?sort=trending&search=blockchain}} offer textual data from news and social media. These datasets typically operate in isolation, lacking a holistic view of how transaction activities, sentiment, and global market trends interact. Currently, there is a clear need for a comprehensive dataset that integrates multi-source data to model these complex interactions.

To address this gap, this paper introduces the {\bf M}ulti-source {\bf M}ulti-level {\bf M}ulti-token {\bf Eth}ereum {\bf Taskforce} (3MEthTaskforce), a machine learning research platform designed for cryptocurrency risk analysis. The platform offers a rich dataset comprising {\bf 303 million transaction records} from {\bf 3,880 tokens} and {\bf 35 million users}, along with textual sentiment data from Reddit (2014–2024) and key market indicators such as 24-hour trading volumes. By integrating these diverse data sources, 3MEthTaskforce enables comprehensive modeling of the relationships between user behavior, market sentiment, and token performance.

Moreover, existing research on price prediction often lacks the use of state-of-the-art (SOTA) deep learning models \cite{sepehri2025cryptomamba}. Thus, in addition to providing this dataset, 3MEthTaskforce defines two core tasks—{\em User Behavior Prediction} and {\em Token Price Prediction}—and evaluates the dataset's performance using SOTA deep learning models to advance cryptocurrency analysis research. The platform provides benchmarks for each task, featuring 6 dynamic GNN-based models for user behavior prediction and 19 time-series models for token price prediction, with systematic experimental results validating the performance of these models.

\begin{figure*}[t]  
    \centering  
    \includegraphics[width=\linewidth]{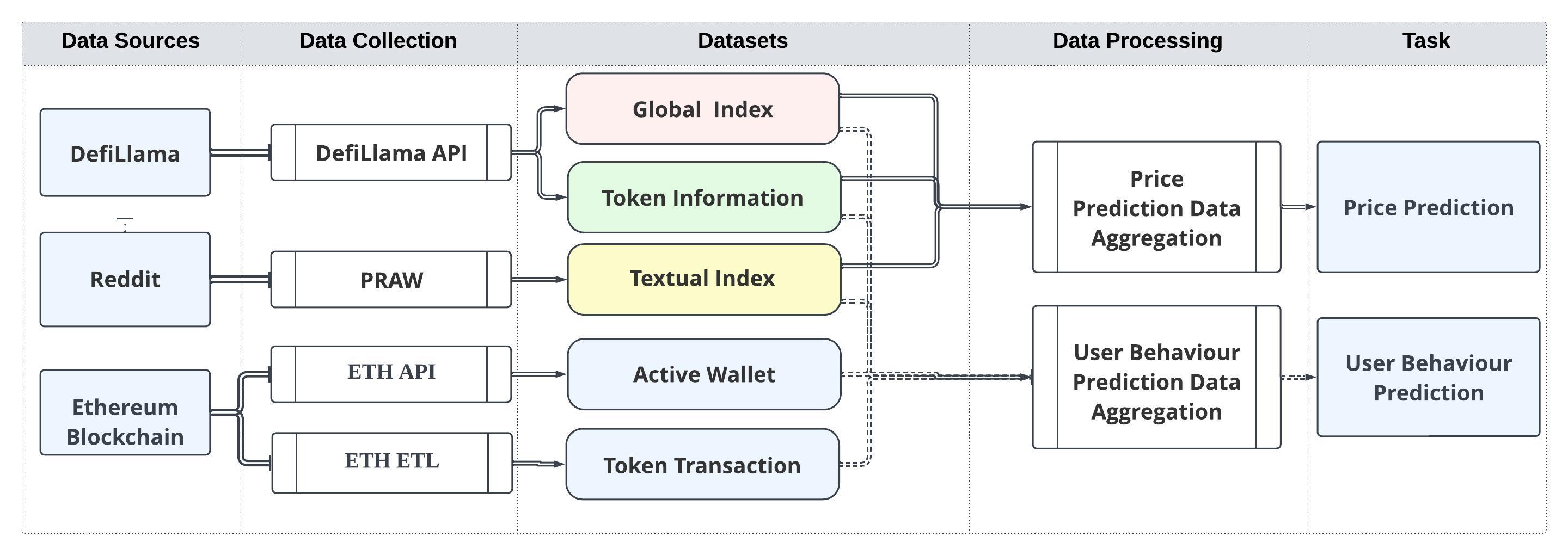}  
    \caption{\footnotesize{3MEthTaskforce Data Pipeline. The third column illustrates datasets: Transaction Records (blue), Token Information (green), Global Market Indices (red), and Textual Indices (yellow).} } 
    \label{fig:Flow Chart}  
\end{figure*}

The key contributions of this paper are summarized as follows:
\begin{itemize}
    \item Introducing 3MEthTaskforce, the first multi-source, multi-token platform that integrates transaction data, token information, textual sentiment data, and market indicators, addressing gaps in existing single-source datasets.  
    \item Defining two critical tasks in cryptocurrency analysis: User Behavior Prediction and Token Price Prediction.  
    \item Offering a comprehensive set of benchmarks for the proposed tasks, validated through systematic experimental results.  
\end{itemize}

\section{Related Work}

\begin{table*}[h!]
\centering
\begin{tabular}{lcccccccc}
\toprule
\multirow{2}{*}{} & \multicolumn{1}{c}{\textbf{Transaction}} & \multicolumn{1}{c}{\textbf{Token}} & \multicolumn{1}{c}{\textbf{Price}} & \multicolumn{1}{c}{\textbf{Global}} & \multicolumn{1}{c}{\textbf{Textual}}  & \multicolumn{1}{c}{\textbf{Active Address}} \\ 
\midrule
TGDEB                   & \checkmark          & 42                 & \(\times\)          & \(\times\)                                    & \(\times\)                & \(\times\)                \\ 
Chartalist                         & \checkmark          & 38                 & \checkmark          & \(\times\)          & \(\times\)                          & \(\times\)                                \\ 
EX-Graph                           & \checkmark          & \(\times\)         & \(\times\)          & \(\times\)          & \checkmark                           & \(\times\)                               \\ 
Public       & \checkmark                   & \(\times\)          & \(\times\)          & \(\times\)                                    & \(\times\)      & \(\times\)                \\ 
CSD  & \(\times\)          & \(\times\)                   & \(\times\)          & \checkmark                                    & \(\times\)      & \(\times\)                \\ 
3MEth                              & \checkmark          & 3880               & \checkmark          & \checkmark          & \checkmark                           & \checkmark                               \\ 
\bottomrule
\end{tabular}
\caption{\footnotesize{The table provides an overview of the datasets and their content. A checkmark (\checkmark) indicates the presence of a specific feature, while a number specifies the exact quantity of tokens included. A ``\(\times\)'' denotes that the corresponding feature is entirely absent from the dataset.}}
\label{tab:datasets}
\end{table*}

\noindent{\bf Cryptocurrency-related data repositories.}
Most existing blockchain data repositories focus on transaction networks. For example, the TGDEB dataset (2021) covers 40 ERC-20 tokens but lacks market indices or user sentiment data \cite{ozturan2021transaction}. Similarly, Chartalist (2022) offers graph learning benchmarks but is limited to a few cryptocurrencies and omits sentiment analysis, global market indices, and user behavior \cite{shamsi2022chartalist}. Public Ethereum datasets on BigQuery, Kaggle, and Athena provide comprehensive transaction data, including complete blockchain records, but lack token prices, global market indices, and social media posts, restricting their use in tasks requiring multi-source data integration. More recent datasets, such as EX-GRAPH (2024), link Ethereum transactions with user social media profiles, enhancing tasks like link prediction and fraud detection, though the text data in this set cannot be used as a market sentiment indicator \cite{wangex}.

Some datasets focus on sentiment in cryptocurrency-related textual data. For example, the Cryptocurrency Sentiment Dataset (CSD) analyzes sentiment in cryptocurrency news and tweets \cite{mohamad2024dataset}, while other similar datasets on Huggingface and Kaggle focus on natural language processing tasks. However, none integrate textual data with price trends, market indices, or user behavior.

In contrast, as shown in Table~\ref{tab:datasets}, the 3MEth dataset integrates multi-token transaction data, long-term community sentiment, and global market indices, offering a comprehensive multimodal dataset for node- and edge-level tasks in blockchain graphs. Its inclusion of community sentiment as a global signal enhances tasks like user behavior analysis and market prediction. 

\smallskip

\noindent {\bf Application of ML in link prediction and price prediction.}
Graph Neural Networks (GNNs) have proven effective for blockchain-related link-based tasks \cite{qi2023blockchain}, including anomaly detection \cite{patel2020graph, han2024mt, hyun2023anti}, user identity inference, and transaction prediction \cite{kim2022graph, liu2020graph, khan2022graph}. However, these studies often overlook key features like token price fluctuations, community sentiment, and market indices. Recent benchmarks such as the Temporal Graph Benchmark and Live Graph Lab \cite{huang2024temporal, zhang2023live} further showcase GNNs' capabilities but highlight the lack of multi-modal data integration in tasks like user classification and link prediction for NFT networks.

Price prediction is typically framed as a time-series problem, with commonly used models including RNN-based models \cite{zoumpekas2020eth, kim2022deep, chen2023analysis, hamayel2021novel}, random forests \cite{zhengyang2019prediction}, and ARIMA \cite{abu2017autoregressive, derbentsev2019forecasting, alahmari2019using, wirawan2019short, yenidougan2018bitcoin}. Although some studies have incorporated social media sentiment \cite{haritha2023cryptocurrency}, they often focus on a limited set of tokens and rely on traditional methods. In contrast, our dataset enables time-series predictions across thousands of cryptocurrencies, incorporating state-of-the-art time-series models for price forecasting, thereby offering broader opportunities for analysis.

\section{The 3MEthTaskforce Dataset}

The 3MEthTaskforce dataset is designed to integrate {\em multi-source}, {\em multi-level}, and {\em multi-token} data. Appendix Raw Data provides some examples of raw data. The dataset offers a rich and diverse collection of data spanning transactions, token information, global market indicators, and social media texts. Below is a detailed description of each section of the datasets. Figure~\ref{fig:Flow Chart} is a flow chart showing the 3MEthTaskforce pipeline from raw data, to datasets, to tasks.

\paragraph{\bf Section 1: Token Transactions} This section provides {\bf 303 million transaction records} from {\bf 3,880 tokens} and {\bf 35 million wallet addresses} on the Ethereum blockchain, stored in {\bf 3,880 CSV files}, each representing a specific token. Each transaction includes:
\begin{itemize}
    \item Blockchain timestamp: Captures transaction timing for temporal analysis.
    \item Sender and receiver wallet addresses: Enables network analysis and user behavior studies.
    \item Token address: Links transactions to tokens for token-specific analysis.
    \item Transaction value: Reflects the number of tokens transferred, essential for liquidity studies.
\end{itemize}
See Table~\ref{tab:token transaction} for examples. Apart from the large csv file, we also provide a smaller csv file containing {\bf $267,242$ transaction records} of {\bf $29,164$  wallet addresses}. This smaller dataset involves a total of $1,194$ tokens, covering the time period $T_s = [\text{Sep 2016}, \text{Nov 2023}]$. This detailed transaction data is critical for studying user behavior, liquidity patterns, and tasks like link prediction and fraud detection.

\paragraph*{\bf Section 2: Token Information} This section offers metadata for {\bf 3,880 tokens}, stored in corresponding CSV files, each with:
\begin{itemize}
    \item Timestamp: Marks the time of data update.
    \item Token price: Useful for price prediction and volatility studies.
    \item Market capitalization: Reflects the token's market size and dominance.
    \item 24-hour trading volume: Indicates the token's liquidity and trading activity.
\end{itemize}
These data support token price prediction and user behavior prediction. By utilizing token addresses from the first section, transactional behaviors can be linked to features such as token prices. This is critical for modeling user-token transactional patterns.

\paragraph*{\bf Section 3: Global Market Indices}
This section provides {\em macro-level data} to contextualize token transactions, stored in separate CSV files. Key indicators include:
\begin{itemize}

    \item Total market capitalization: Measures the overall market's value, with breakdowns by token type.
    \item Bitcoin dominance: Tracks Bitcoin's share of the cryptocurrency market.
    \item Stablecoin market capitalization: Highlights stablecoin liquidity and stability.
    \item 24-hour trading volume: A key measure of market activity.
\end{itemize} 
These metrics are crucial for incorporating global market trends into volatility forecasting and user behavior models.

\begin{figure*}
    \centering
    \includegraphics[width=1\linewidth]{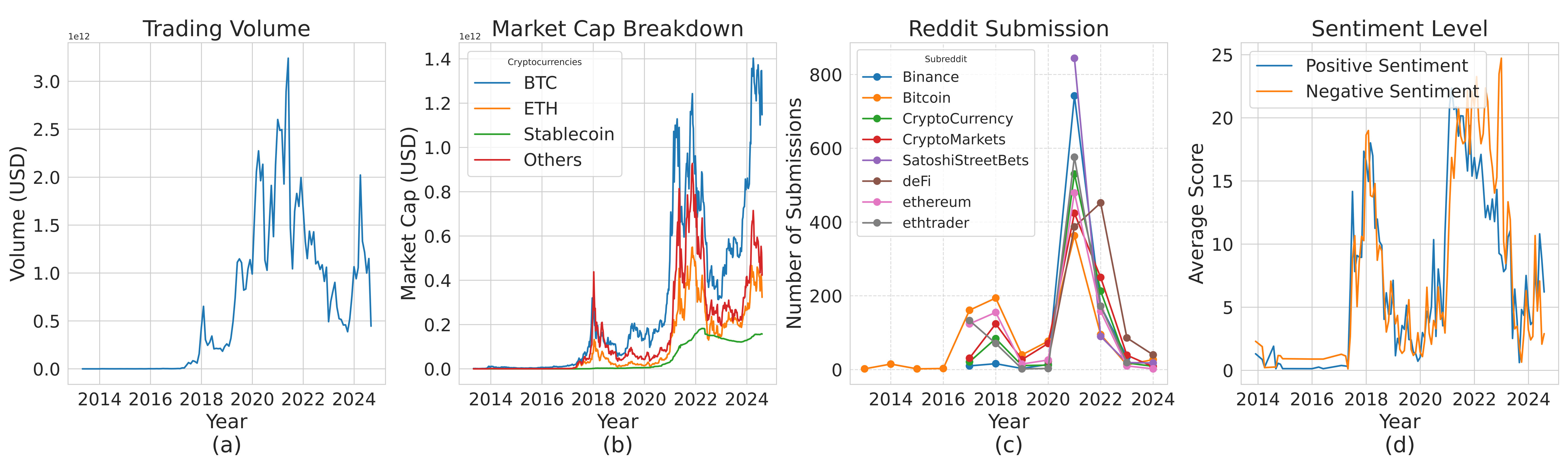}
    \caption{\footnotesize{Time-series analysis of cryptocurrency market activity and online discourse from data in this dataset. \textbf{(a)} displays the aggregate trading volume, indicating periods of high market activity. \textbf{(b)} shows the market capitalization breakdown by cryptocurrency category, revealing the relative dominance of BTC over time. \textbf{(c)} illustrates the temporal dynamics of submissions across various cryptocurrency subreddits. \textbf{(d)} presents the average positive and negative sentiment scores, revealing fluctuations in community perception of the cryptocurrency market.}}
    \label{fig:enter-label}
\end{figure*}

\paragraph*{\bf Section 4: Textual Indices}
This section contains sentiment data from Reddit's Ethereum community, covering {\bf 7,800 top posts from 2014 to 2024}. Each post includes:
\begin{itemize}
    \item Timestamp: Aligns sentiment with price movements.
    \item Comment text: For sentiment analysis and NLP tasks.
    \item Post score (net upvotes): Reflects engagement and sentiment strength.
    \item Number of comments: Gauges sentiment intensity.
    \item Sentiment indices: Sentiment scores computed using methods presented in Section~\ref{sec:Data preprocess}. 
\end{itemize}

This data is valuable for understanding social dynamics in the market and enhancing sentiment analysis models that can explain market movements and improve behavioral predictions.

\subsection{Data Collection}
We employed a multi-source strategy, integrating data from three providers to ensure diversity and representativeness. All the tools we utilized are open-source, and according to their licensing terms—PRAW\footnote{\url{https://github.com/praw-dev/praw?tab=BSD-2-Clause-1-ov-file}}, DefiLlama\footnote{\url{https://github.com/orgs/DefiLlama/repositories?type=all}}, and Ethereum Public ETL\footnote{\url{https://github.com/blockchain-etl/ethereum-etl?tab=MIT-1-ov-file}}—redistribution is permitted as long as proper attribution is provided. 

\noindent {\bf Ethereum Public ETL Tool for Token Transactions:} We used the Ethereum Public ETL tool \cite{medvedev2018d5} to collect token transaction data for the 3MEth dataset. This tool efficiently extracts, transforms, and loads transaction data from the Ethereum blockchain, providing $3,880$ token transaction records. Additionally, we used a free Ethereum API (\url{etherscan.io/}) to gather $5,855$ active wallet addresses and their transactions. This results in $267,242$ transaction records among $29,164$  wallet addresses and $1,194$ tokens which form our smaller dataset. 

\noindent {\bf DefiLlama for General Token Information and Global Index:} DefiLlama\footnote{\url{https://defillama.com/}} was used to gather historical prices for $3,880$ tokens and global market indices, accessing data from multiple exchanges.

\noindent {\bf Reddit API (PRAW) for Textual Data:} The Python Reddit API Wrapper (PRAW)\footnote{\url{https://praw.readthedocs.io/en/stable/}} was used to extract approximately $7,800$ timestamped posts from eight popular cryptocurrency subreddits (e.g., r/Ethereum, r/CryptoCurrency).

\subsection{Dataset Statistics}

The transaction records and historical price data of tokens are vast and complex, but global market indicators provide a rough overview of their trends and interrelationships. Trading volume reflects the interaction between transaction frequency and price, while market capitalization captures the product of price and circulating supply. Social media sentiment further enriches the understanding of market dynamics.

As shown in Figure \ref{fig:enter-label}, the trading volume (a) exhibited a sharp increase from 2014 to 2024, peaking at over 3 trillion USD between 2020 and 2022, before gradually declining. The market capitalization breakdown (b) shows BTC dominance throughout the period, with ETH and stablecoins gaining market share after 2020, correlating with the peak trading volume.

Reddit submissions (c) mirrored this trend, with a surge in activity across cryptocurrency subreddits between 2020 and 2021. Sentiment scores (d) fluctuated in tandem with trading volume, reflecting the direct influence of market dynamics on community sentiment.

These figures collectively highlight consistent time-series trends, demonstrating how market activity, community engagement, and sentiment interplay to shape the cryptocurrency ecosystem.

\subsection{Ethics, Privacy, IP and FAIR Data Principles}

This research adheres to ethical guidelines to minimize privacy risks(Appendix Datasheets for Datasets). Ethereum's public blockchain and publicly sourced Reddit data ensure transparency, with no private content accessed(Appendix Ethics and Privacy). . The dataset respects the intellectual property(IP) rights of Reddit content by providing only sentiment scores labeled by a large language model, without including the original text. We offer metadata retrieval code, allowing readers to apply for API access to obtain the original data themselves. The study aims to provide valuable insights into cryptocurrency trading and decentralized finance, with potential benefits outweighing minimal risks. The dataset respects the intellectual property rights of Reddit content by providing only sentiment scores labeled by a large language model, without including the original text. 
\paragraph{FAIR.}
The dataset provided in this study adheres to the FAIR principles \cite{gebru2021datasheets}, exhibiting the following characteristics:

\textbf{Findable:} The dataset is publicly available through the figshare platform and has been assigned a permanent Digital Object Identifier (DOI) (https://figshare.com/s/42e1bec80cd0b1809d14).

\textbf{Accessible:} The dataset is freely accessible on the Internet, allowing any user with an Internet connection to retrieve it. All data are provided in the comma-separated value (CSV) format, which is a standard format for managing tabular data.

\textbf{Interoperable:} The dataset can be easily loaded and viewed using most modern database management systems or spreadsheet software.

\textbf{Re-usable:} The release page includes a comprehensive README file for quick reference and a standardised metadata in JSON format facilitating re-use.

\section{Tasks}

To validate the potential applications of the dataset, the 3MEthTaskforce platform introduces two key tasks: \textit{User Behavior Prediction} and \textit{Token Price Prediction}. These tasks utilize the platform's diverse dataset to model and predict interactions between users and tokens, providing a framework to systematically study.

\textbf{We assume that each wallet address represents a unique user.} To formally define these tasks, we introduce the following key terminologies:

\begin{itemize}
    \item \textbf{Transaction History}: Captures all user buy and sell activities across various tokens over a time period \( (1, 2, \dots, \tau) \), provided by the Token Transactions section (Section 1). For a user \( i \in U \), token \( j \in C \), and time step \( t \leq \tau \), let \( a_{i,j,t} \in \{0, 1\} \) denote a transaction, where \( a_{i,j,t} = 1 \) indicates a buy/sell transaction. The transaction matrix is:
    \[
    \bm{A} = (a_{i,j,t})_{i \in U, j \in C, t \leq \tau}.
    \]

    \item \textbf{Token-Level Features}: Provided by the Token Information section (Section 2), these include time-dependent features for each token \( j \in C \):
    \begin{itemize}
        \item \textit{Price}: \( p_{j,t} \), the market value of token \( j \) at time \( t \).
        \item \textit{Market Capitalization}: \( m_{j,t} \), calculated as the product of the token price and circulating supply.
        \item \textit{Volume}: \( v_{j,t} \), the trading volume at time \( t \).
    \end{itemize}
    These features are represented as time-series vectors over \( 1, \dots, \tau \):
    \[
    \bm{p}_j = (p_{j,1}, \dots, p_{j,\tau}),
    \]
    \[
    \bm{m}_j = (m_{j,1}, \dots, m_{j,\tau}),
    \]
    \[
    \bm{v}_j = (v_{j,1}, \dots, v_{j,\tau}).
    \]

    \item \textbf{Global Market Indices}: Provided by the Global Market Indices section (Section 3), these reflect broader market trends, such as Bitcoin dominance and total market capitalization. The time-series vector is:
    \[
    \bm{g} = (g_1, g_2, \dots, g_\tau).
    \]

    \item \textbf{Market Sentiment}: Derived from the Textual Indices section of the dataset, market sentiment is captured by a sentiment index \( s_t \), representing collective emotions and opinions. The sentiment is represented as a time-series vector:
    \[
    \bm{s} = (s_1, s_2, \dots, s_\tau).
    \]
\end{itemize}

This provides the essential terminologies for modeling user behavior and market conditions in our tasks.

\subsection{User behavior prediction}

This task aims to forecast users' purchasing and selling behaviors in the cryptocurrency market. By predicting which users are likely to buy or sell tokens and identifying the specific tokens they are most likely to transact with, stakeholders -- such as traders, investors, and institutions -- can anticipate sharp market movements and respond proactively. 
For this task, we consider the interplay of multiple factors influencing user behavior.

\paragraph*{\bf Task definition.} The goal of the {\em user behavior prediction} task is to forecast future transactions \( \bm{A}_{:,:,t} \) at a future time \( t > \tau \), based on historical data. The task can be formally defined as follows:

\begin{table}[h]
\centering
\begin{tabular}{p{1cm}p{6cm}}
\hline
\multicolumn{2}{l}{\bf User behavior prediction task} \\
\hline 
{\bf Input:}   & a transaction matrix $\bm{A}$, price vectors $(\bm{p}_j)_{j\in C}$, market capitalization vectors $(\bm{m}_j)_{j\in C}$ and volume vectors $(\bm{v}_j)_{j\in C}$ of all tokens, global market vector $\bm{g}$ and market sentiment vector $\bm{s}$. \\
{\bf Output:}   & predicted transactions $\bm{A}_{:,:,t}$ at time $t> \tau$  \\ \hline
\end{tabular}
\end{table}

\subsection{Token Price Prediction}

This task involves forecasting the future price of a cryptocurrency token, providing insights into price fluctuations and supporting informed investment decisions.
We approach this task in a broader market context by incorporating historical token data, global market indices, and sentiment analysis from the textual indices. The task is formulated as:

\begin{table}[h]
\centering
\begin{tabular}{p{1cm}p{6cm}}
\hline
\multicolumn{2}{l}{\bf Token price prediction task} \\
\hline 
{\bf Input:}   & price vectors $(\bm{p}_j)_{j\in C}$ of all tokens, global market vector $\bm{g}$ and market sentiment vector $\bm{s}$. \\
{\bf Output:}    & predicted prices of all tokens $(p_{j,t})_{j\in C}$ at time $t > \tau$  \\ \hline
\end{tabular}
\end{table}


\section{User behavior prediction}
\label{sec:Data preprocess}

\subsection{Baseline methods}
We treat this task as a {\em link prediction task} on an {\em edge-labelled temporal bipartite graph}(see details in Appendix User Behaviour Experiment Setup). We compare GNN models, input features, and sentiment features to predict user behavior. 

\paragraph*{\bf Graph Definition.}  
The graph comprises vertices \( V = U \cup C \), where \( U \) represents users and \( C \) represents tokens. A temporal edge is defined as \( (u, c_j, t) \), where \( u \in U \), \( c_j \in C \), and \( 1 \leq t \leq \tau \), indicating a transaction at timestamp \( t \). Each edge \( (u, c_j, t) \) is associated with \( q \) (transaction amount) and \( \mathrm{trans}_{j,t} \) (transaction label), forming a labelled temporal edge \( (u, c_j, q, \mathrm{trans}_{j,t}) \). The graph \( (V, E) \) includes all such edges up to timestamp \( \tau \). The goal is to predict whether a temporal edge \( (u, c, t') \) will appear for \( t' > \tau \).

\paragraph*{\bf Sentiment Index.}  
Sentiment indices are derived from top posts at time \( t \), concatenated into \( \mathbf{txt}_t \). The prompts used for these methods are provided in Appendix Prompt. The sentiment index aggregation methods are provided in Appendix LLM Knowledge Base Marked Sentiment
Data Algorithm. Two methods are used:
\begin{enumerate}
    \item Overall sentiment score \( s^{\text{overall}}_t \) \cite{bhatt2023machine}, rated 0--10 (5 is neutral).
    \item Separate positive and negative scores \( s^{\text{pos}}_t \) and \( s^{\text{neg}}_t \), incorporating temporal context.
\end{enumerate}

\subsection{User Behavior Experiment Setup}

\paragraph*{\bf Dataset and Settings.}  Five transaction input vectors are defined:
\begin{itemize}
    \item \( \text{trans\_record}_{j,t} = (p_{j,t}, m_{j,t}, v_{j,t}) \): token information.
    \item \( \text{trans\_global}_{j,t} = g_t \): global market index.
    \item \( \text{trans\_text}_{j,t} = s^{\text{overall}}_t \): overall sentiment score.
    \item \( \text{trans\_text\_llm}_{j,t} = (s^{\text{pos}}_t, s^{\text{neg}}_t) \): positive and negative scores.
    \item \( \text{trans\_all}_{j,t} = (p_{j,t}, m_{j,t}, v_{j,t}, g_t, s^{\text{overall}}_t, s^{\text{pos}}_t, s^{\text{neg}}_t) \): all features.
\end{itemize}

We use a subset with $260,000$ transactions and $29,164$ wallet addresses, split 70\%/15\%/15\% for train/validation/test. Models are trained using Adam \cite{diederik2014adam} with binary cross-entropy loss, a learning rate of 0.0001, batch size 200, and early stopping after 20 epochs of no improvement.

\paragraph*{\bf Evaluation Metrics.}  
Performance is evaluated using test set average precision (TAP) and new node average precision (NAP) \cite{xu2020inductive, poursafaei2022towards}. Results are averaged over three runs.
3MEthTaskforce implements {\bf six baseline dynamic GNN models} to the constructed graph $G$: DyGFormer  \cite{yu2023towards}, JODIE  \cite{kumar2019predicting}, DyRep  \cite{trivedi2019dyrep}, TGAT  \cite{xu2020inductive}, TGN  \cite{rossi2020temporal}, and TCL  \cite{wang2021tcl}. These models were chosen as they were recently proposed for this task and have demonstrated promising performance in various link prediction tasks. For each of these models, we measure the performance of link prediction using all six transaction label vectors above.  A detailed description of these models can be found in Appendix GNN Model.

\subsection{Baseline Evaluation Results}
The confidence level was set to 0.05, with results averaged over three experiments and errors within the confidence interval. Table~\ref{tab:GNN results} reports the performance of six models across datasets using TAP and NAP as metrics.

\begin{table*}[ht]
\centering

\begin{tabular}{lcccccccccccc}
\toprule
\multirow{2}{*}{\textbf{Dataset}} & \multicolumn{2}{c}{\textbf{DyRep}} & \multicolumn{2}{c}{\textbf{TCL}} & \multicolumn{2}{c}{\textbf{TGN}} & \multicolumn{2}{c}{\textbf{TGAT}} & \multicolumn{2}{c}{\textbf{JODIE}} & \multicolumn{2}{c}{\textbf{DyGFormer}} \\ 
\cmidrule(r){2-13}
 & \textbf{tap} & \textbf{nap} & \textbf{tap} & \textbf{nap} & \textbf{tap} & \textbf{nap} & \textbf{tap} & \textbf{nap} & \textbf{tap} & \textbf{nap} & \textbf{tap} & \textbf{nap} \\ 
\midrule
trans & 0.914 & 0.870 & 0.859 & 0.778 & 0.917 & 0.860 & \underline{0.878} & \underline{0.794} & \textbf{0.935} &\textbf{ 0.898} & 0.928 & 0.879 \\
trans\_record & 0.918 & 0.876 & \underline{0.870} & \underline{0.794} & 0.899 & 0.834 & 0.861 & 0.773 & \textbf{0.930} & \textbf{0.890} & 0.925 & 0.879 \\
trans\_global & 0.909 & 0.864 & 0.853 & 0.767 & 0.881 & 0.810 & 0.834 & 0.750 & 0.931 & 0.889 & \underline{\textbf{0.939}} & \underline{\textbf{0.905}} \\
trans\_text & 0.923 & 0.876 & 0.835 & 0.737 & 0.917 & 0.861 & 0.826 & 0.705 & \textbf{0.942} & \textbf{0.900} & 0.919 & 0.870 \\
trans\_text\_llm & \underline{0.926} & \underline{0.881} & 0.867 & 0.786 & \underline{0.926} & \underline{0.873} & 0.841 & 0.733 & \underline{\textbf{0.955}} & \underline{\textbf{0.920}} & 0.938 & 0.896 \\
trans\_all & 0.922 & 0.869 & 0.846 & 0.750 & 0.887 & 0.806 & 0.845 & 0.743 & \textbf{0.941} & \textbf{0.898} & 0.913 & 0.860 \\
\bottomrule
\end{tabular}
\caption{ %
    \footnotesize{Comparison of six dynamic GNN models on six datasets for the User Prediction task. TAP is the average precision on the test set, and NAP is the average precision on the new node set. An \underline{underline} indicates the best result in each column, and \textbf{bold} font highlights the best result in each row.%
}}
\label{tab:GNN results}
\end{table*}

\paragraph{Impact of LLM-enhanced sentiment}
Models incorporating LLM-derived sentiment scores (\texttt{trans\_text\_llm}) outperform those using only overall sentiment (\texttt{trans\_text}). For example, TCL improves from 0.737 NAP on \texttt{trans\_text} to 0.786 on \texttt{trans\_text\_llm}, demonstrating the benefit of incorporating background knowledge from LLMs for sentiment annotation.

\paragraph{Effectiveness of additional features}
Adding features like token information and sentiment improves performance across models. For instance, DyRep achieves a TAP of 0.926 on \texttt{trans\_text\_llm}, compared to 0.914 using only transaction data. Similarly, DyGFormer sees a TAP increase from 0.928 on transactions to 0.938 on \texttt{trans\_text\_llm}, indicating that these additional features help predict user-token interactions more effectively.

\paragraph{Performance of different GNN models}
JODIE and DyGFormer consistently perform best, with JODIE reaching a TAP of 0.941 on \texttt{trans\_all} and 0.955 on \texttt{trans\_textual\_llm}. DyRep shows stable results, with a high of 0.926 on \texttt{trans\_text\_llm}, while TCL performs weaker, reaching only 0.750 TAP on \texttt{trans\_all}.

\section{Price Prediction}

\sisetup{scientific-notation=true}

\begin{table*}[h!]
\centering
\resizebox{\textwidth}{!}{%
\begin{tabular}{lccccccccc}
\toprule
\multirow{2}{*}{\textbf{Model Name}}& \multicolumn{2}{c}{\textbf{LUNC}} & \multicolumn{2}{c}{\textbf{BTC}} & \multicolumn{2}{c}{\textbf{ETH}} & \multirow{2}{*}{\textbf{Rank}} \\
\cmidrule(r){2-3} \cmidrule(r){4-5} \cmidrule(r){6-7}
 & \textbf{mse} & \textbf{mae} & \textbf{mse} & \textbf{mae} & \textbf{mse} & \textbf{mae} & \\
\midrule
Crossformer & \num{7.79e-1} & \num{8.37e+0} & \num{3.35e+2} & \num{1.65e+2} & \num{9.9e+0}  & \num{2.26e+1} & 15.5 \\
DLinear     & \num{1.67e-3} & \num{3.25e-1} & \num{6.8e+1}  & \num{5.99e+1} & \num{1.02e+1} & \num{2.36e+1} & 11.67 \\
FEDformer   & \num{3.91e-1} & \num{4.9e+0}  & \num{3.93e+1} & \num{4.35e+1} & \num{1.19e+1} & \num{2.44e+1} & 13.33 \\
FiLM        & \num{1.97e-5} & \num{4.43e-2} & \num{3.1e+1}  & \num{3.89e+1} & \num{8.28e+0} & \num{1.97e+1} & 5 \\
FITS        & \num{1.04e-7} & \num{2.64e-3} & \num{3.12e+1} & \num{4.12e+1} & \underline{\num{6.68e+0}} & \num{1.89e+1} & \underline{4.5} \\
Informer    & \num{3.95e-4} & \num{1.60e-1} & \num{4.52e+1} & \num{4.78e+1} & \num{1.27e+1} & \num{2.44e+1} & 11 \\
iTransformer & \num{2.83e-3} & \num{4.89e-1} & \num{3.49e+1} & \num{4.27e+1} & \num{9.61e+0} & \num{2.10e+1} & 9.83 \\
MICN        & \num{5.98e-2} & \num{1.94e+0} & \num{1.11e+2} & \num{7.68e+1} & \num{1.52e+1} & \num{2.92e+1} & 16 \\
NLinear     & \num{3.30e-1} & \num{4.63e+0} & \num{2.28e+1} & \num{2.19e+1} & \num{6.96e+0} & \num{1.79e+1} & 7 \\
Nonstationary\_Transformer & \num{2.67e-3} & \num{5.14e-1} & \num{3.22e+1} & \num{3.95e+1} & \num{1.20e+1} & \num{2.45e+1} & 11.33 \\
PatchTST    & \num{2.40e-3} & \num{4.85e-1} & \num{2.65e+1} & \num{3.51e+1} & \num{7.75e+0} & \num{1.84e+1} & 7.75 \\
LinearRegression & \underline{\num{7.44e-11}} & \underline{\num{6.48e-5}} & \num{2.31e+1} & \num{3.14e+1} & \num{7.06e+0} & \num{1.77e+1} & 8 \\
RNN         & \num{4.24e-1} & \num{6.21e+0} & \num{7.07e+1} & \num{5.15e+1} & \num{1.40e+1} & \num{3.00e+1} & 16 \\
TCN         & \num{1.41e-4} & \num{1.19e-1} & \num{7.00e+2} & \num{1.12e+2} & \num{3.07e+1} & \num{4.62e+1} & 14.5 \\
TimeMixer   & \num{1.09e-4} & \num{8.35e-2} & \num{4.64e+1} & \num{4.51e+1} & \num{1.01e+1} & \num{2.23e+1} & 8.5 \\
TimesNet    & \num{1.37e-3} & \num{3.23e-1} & \num{2.86e+1} & \num{1.01e+1} & \num{8.93e+0} & \num{1.96e+1} & 6.83 \\
Triformer   & \num{4.14e-3} & \num{1.24e+0} & \num{9.64e+1} & \num{9.68e+1} & \num{9.64e+0} & \num{2.03e+1} & 13 \\
ARIMA       & \num{2.86e-8} & \num{1.46e-3} & \underline{\num{1.45e+0}} & \num{4.53e+1} & \num{7.75e+0} & \num{1.79e+1} & 10 \\
TimeGPT     & \num{2.00e-4} & \num{1.00e-1} & \num{5.23e+3} & \underline{\num{6.41e+0}} & \num{5.77e+3} & \underline{\num{9.01e+0}} & 11.17 \\
\bottomrule
\end{tabular}%
}
\caption{\footnotesize{The univariate forecasting results indicate that ARIMA and the regression model consistently outperformed other models in forecasting tasks for LUNC, BTC, and ETH. FITS demonstrated strong overall performance, while TimeGPT performed the worst across all datasets. \underline{Underlined values} in the table indicate the best-performing results.
}}
\label{tab:forc_results}
\end{table*}

\begin{table*}[h!]

\centering
\resizebox{\textwidth}{!}{%
\begin{tabular}{lccccccccc}
\toprule
\multirow{2}{*}{\textbf{Model Name}} & \multicolumn{2}{c}{\textbf{Price}} & \multicolumn{2}{c}{\textbf{Global}} & \multicolumn{2}{c}{\textbf{Textual}} & \multicolumn{2}{c}{\textbf{All}} & \multirow{2}{*}{\textbf{Rank}} \\
\cmidrule(lr){2-3} \cmidrule(lr){4-5} \cmidrule(lr){6-7} \cmidrule(lr){8-9}
 & \textbf{wape} & \textbf{msmape} & \textbf{wape} & \textbf{msmape} & \textbf{wape} & \textbf{msmape} & \textbf{wape} & \textbf{msmape} & \\
\midrule
Crossformer & 36.73 & 97.74 & 38.57 & 102.59 & 38.14 & 102.07 & 36.46 & 74.36 & 17 \\
DLinear & 19.67 & 26.44 & 16.88 & 24.98 & 21.11 & 28.11 & 18.02 & 26.59 & 10.25 \\
FEDformer & 27.33 & 66.74 & 24.05 & 61.88 & 28.4 & 67.22 & 24.06 & 54.81 & 15.12 \\
FiLM & 17.54 & 17.18 & 15.22 & 16.71 & 18.97 & \underline{18.72} & 16.32 & 18.16 & \underline{3.62} \\
FITS & 18.31 & 17.85 & 16.08 & 17.46 & 19.69 & 19.3 & 17.14 & 18.81 & 5.62 \\
Informer & 21.18 & 20.86 & 20.21 & 21.25 & 21.77 & 20.87 & 19.23 & 21.05 & 10.25 \\
iTransformer & \underline{17.01} & \textbf{16.23} & 14.69 & \textbf{15.68} & \textbf{18.35} & \textbf{17.64} & \underline{15.76} & \textbf{17.03} & \textbf{1.5} \\
MICN & 22.53 & 34.27 & 20.45 & 36.18 & 24.33 & 35.02 & 20.07 & 33.31 & 13.62 \\
NLinear & 17.42 & 20.91 & 14.78 & 19.06 & \underline{18.79} & 22.32 & 15.85 & 20.4 & 5.88 \\
Nonstationary\_Transformer & 19.8 & 19.47 & 16.97 & 19 & 20.47 & 20.21 & 17.52 & 19.7 & 7.62 \\
PatchTST & 17.62 & \underline{16.48} & 15.44 & \underline{16.08} & 19.05 & 18.08 & 16.52 & \underline{17.55} & 3.75 \\
RNN & 54.57 & 78.55 & 47.11 & 72.64 & 48.63 & 70.1 & 45.27 & 56.43 & 17 \\
TimeMixer & 18.6 & 18.52 & 15.32 & 17.54 & 19.31 & 19.08 & 16.2 & 18.69 & 5 \\
TimesNet & 21.41 & 21.5 & 16.33 & 18.38 & 21.69 & 21.16 & 18.14 & 20.15 & 9.25 \\
Triformer & 24.97 & 30.84 & 23.42 & 32 & 28.81 & 36.02 & 24.8 & 31.45 & 14.12 \\
LinearRegression & 36.36 & 30.41 & \textbf{10.79} & 26.54 & 199.39 & 34.45 & \textbf{13.17} & 29.53 & 10.5 \\
TCN & \textbf{14.89} & 30.7 & \underline{11.19} & 189.98 & 50.98 & 37.47 & 17.6 & 190.13 & 11.62 \\
\bottomrule
\end{tabular}%
}
\caption{\footnotesize{The multivariate forecasting performance improved significantly with the inclusion of additional features compared to using only price data. This highlights the advantage of incorporating diverse data features in enhancing model performance for price forecasting tasks. \textbf{Bold values} represent the best performance in each category, while \underline{underlined values} indicate the second-best performance.
}}
\label{tab:datasets_results}

\end{table*}

\subsection{Baseline Methods}

We approach the price prediction task as a problem of \emph{time series forecasting} (see details in Appendix Price Prediction Experiment Setup). To evaluate the performance of various forecasting methods on the dataset, we incorporate 19 methods, which can be broadly categorized into three groups: statistical learning, machine learning, and deep learning. For further details, see Appendix Time Series Forecasting Model Descriptions.

In the category of statistical learning, we include ARIMA \cite{box1970distribution}. For machine learning methods, we introduce Linear Regression \cite{zou2003correlation}. Lastly, deep learning methods are further subdivided into the following categories:

\begin{itemize}
    \item \textbf{RNN-based Models}: RNN \cite{sherstinsky2020fundamentals}
    \item \textbf{CNN-based Models}: MICN \cite{wang2023micn}, TimesNet \cite{wu2022timesnet}, and TCN \cite{hewage2020temporal}
    \item \textbf{MLP-based Models}: TimeMixer \cite{wang2024timemixer}, NLinear \cite{zeng2023transformers}, and DLinear \cite{qiu2024tfb}
    \item \textbf{Transformer-based Models}: PatchTST \cite{nie2022time}, FITS \cite{xu2023fits}, Crossformer \cite{zhang2023crossformer}, FEDformer \cite{zhou2022fedformer}, iTransformer \cite{liu2023itransformer}, Non-stationary Transformer \cite{liu2022non}, Informer \cite{zhou2021informer}, and Triformer \cite{cirstea2022triformer}
    \item \textbf{Model-Agnostic Models}: FiLM \cite{zhou2022film}
    \item \textbf{LLM-based Models}: TimeGPT \cite{garza2023timegpt}
\end{itemize}

\subsection{Price Prediction Experiment Setup}

The experiment aims to (1) compare time series models for predicting single and multiple token prices, (2) evaluate the impact of global indices and sentiment features, and (3) assess model performance based on token listing time. The task is divided into two parts:

\paragraph{Univariate Forecasting.}  
For three tokens (LUNC, BTC, ETH), the input vectors are:  
\begin{itemize}
    \item \( \text{price\_lunc}_{j,t} = (p^{\text{lunc}}_{j,t}) \)
    \item \( \text{price\_btc}_{j,t} = (p^{\text{btc}}_{j,t}) \)
    \item \( \text{price\_eth}_{j,t} = (p^{\text{eth}}_{j,t}) \)
\end{itemize}

\paragraph{Multivariate Forecasting.}  
For 211 tokens with at least six years of data, the input vectors are:  
\begin{itemize}
    \item \( \text{price}_{j,t} = p_{j,t} \)
    \item \( \text{price\_global}_{j,t} = (p_{j,t}, g_{j,t}) \)
    \item \( \text{price\_text}_{j,t} = (p_{j,t}, s^{\text{pos}}_t, s^{\text{neg}}_t) \)
    \item \( \text{price\_all}_{j,t} = (p_{j,t}, g_{j,t}, s^{\text{pos}}_t, s^{\text{neg}}_t) \)
\end{itemize}

\paragraph{Parameter Settings.}  
For univariate forecasting, we use a fixed forecasting horizon of 14 and a look-back window of 1.25 times the horizon \cite{liu2022non, qiu2024tfb}. For multivariate forecasting, we adopt a rolling horizon of 60 \cite{qiu2024tfb}. Models are trained with the L2 loss function and ADAM optimizer \cite{diederik2014adam} using PyTorch \cite{paszke2019pytorch}. The batch size is initially set to 32 but can be halved to a minimum of 8 if memory issues occur.

\paragraph{Performance Metrics.}  
Metrics for univariate forecasting include Normalized Mean Squared Error (MSE) and Normalized Mean Absolute Error (MAE). For multivariate forecasting, Normalized Weighted Absolute Percentage Error (WAPE) and Mean Symmetric Mean Absolute Percentage Error (MSMAPE) are used, ensuring comparability across datasets by eliminating scale differences.

\subsection{Baseline Evaluation Results}

All results were averaged over three experiments with a confidence level of 0.05, and errors fell within the confidence interval.

\subsubsection{Univariate Forecasting.}  
ARIMA and the regression model outperformed other models for LUNC, BTC, and ETH (Table~\ref{tab:forc_results}). The regression model achieved the lowest MSE ($7.44 \times 10^{-11}$) and MAE ($6.48 \times 10^{-5}$) for LUNC. ARIMA yielded the best results for BTC (MSE: 1.45, MAE: 1.46) and ETH (MSE: 1.45, MAE: 1.79). FITS also performed well, ranking 4.5 overall, while TimeGPT showed the poorest performance, with MSE values exceeding $5 \times 10^3$ for BTC and ETH.

\subsubsection{Multivariate Forecasting.}  
iTransformer demonstrated the best performance across datasets, ranking 1.5 overall, with the lowest msMAPE on the Global (15.68) and All (17.03) datasets (Table~\ref{tab:datasets_results}). TCN performed well on the Price dataset, achieving the lowest wAPE (14.89) and msMAPE (11.19). In contrast, RNN performed the worst, with a wAPE of 54.57 and msMAPE of 78.55 on the Price dataset, and similar poor results on the Global dataset. 

\textbf{Feature Impact.} Integrating additional features significantly improved forecasting accuracy. For instance, iTransformer's msMAPE on the All dataset (17.03) was notably better than on the Price dataset (16.23), highlighting the value of diverse data features.

\textbf{Task Comparison.} Models such as iTransformer excelled in multivariate forecasting but showed variable performance in univariate tasks, with an average ranking of 1.5 in multivariate and 11 in univariate forecasting.

\section{Limitations and Broader Impact}

The 3MEthTaskforce platform faces limitations such as incomplete data coverage, potential biases in sentiment data from platforms like Reddit, and untested generalization across broader datasets and markets. Privacy risks, despite anonymization, also persist. However, its multimodal dataset supports comprehensive analysis of user behavior and market dynamics, offering valuable benchmarks and advancing decentralized finance research, making it a key tool for exploring cryptocurrency risks and driving blockchain innovation.

\bigskip
\noindent

\bibliography{aaai25}

\begin{thebibliography}{65}
\providecommand{\natexlab}[1]{#1}

\bibitem[{Abu~Bakar and Rosbi(2017)}]{abu2017autoregressive}
Abu~Bakar, N.; and Rosbi, S. 2017.
\newblock Autoregressive integrated moving average (ARIMA) model for
  forecasting cryptocurrency exchange rate in high volatility environment: A
  new insight of bitcoin transaction.
\newblock \emph{International Journal of Advanced Engineering Research and
  Science}, 4(11): 130--137.

\bibitem[{Akila et~al.(2023)Akila, Nitin, Prasanth, Reddy, and
  Kumar}]{akila2023cryptocurrency}
Akila, V.; Nitin, M.; Prasanth, I.; Reddy, S.; and Kumar, A. 2023.
\newblock A Cryptocurrency Price Prediction Model using Deep Learning.
\newblock In \emph{E3S Web of Conferences}, volume 391, 01112. EDP Sciences.

\bibitem[{Alahmari(2019)}]{alahmari2019using}
Alahmari, S.~A. 2019.
\newblock Using Machine Learning ARIMA to Predict the Price of
  Cryptocurrencies.
\newblock \emph{ISeCure}, 11(3).

\bibitem[{Bhatt, Ghazanfar, and Amirhosseini(2023)}]{bhatt2023machine}
Bhatt, S.; Ghazanfar, M.; and Amirhosseini, M. 2023.
\newblock Machine learning based cryptocurrency price prediction using
  historical data and social media sentiment.
\newblock \emph{Computer Science \& Information Technology (CS \& IT)}, 13(10):
  1--11.

\bibitem[{Box and Pierce(1970)}]{box1970distribution}
Box, G.~E.; and Pierce, D.~A. 1970.
\newblock Distribution of residual autocorrelations in
  autoregressive-integrated moving average time series models.
\newblock \emph{Journal of the American statistical Association}, 65(332):
  1509--1526.

\bibitem[{Briola et~al.(2023)Briola, Vidal-Tom{\'a}s, Wang, and
  Aste}]{briola2023anatomy}
Briola, A.; Vidal-Tom{\'a}s, D.; Wang, Y.; and Aste, T. 2023.
\newblock Anatomy of a Stablecoin’s failure: The Terra-Luna case.
\newblock \emph{Finance Research Letters}, 51: 103358.

\bibitem[{Chen(2023)}]{chen2023analysis}
Chen, J. 2023.
\newblock Analysis of bitcoin price prediction using machine learning.
\newblock \emph{Journal of Risk and Financial Management}, 16(1): 51.

\bibitem[{Chen, Li, and Sun(2020)}]{chen2020bitcoin}
Chen, Z.; Li, C.; and Sun, W. 2020.
\newblock Bitcoin price prediction using machine learning: An approach to
  sample dimension engineering.
\newblock \emph{Journal of Computational and Applied Mathematics}, 365: 112395.

\bibitem[{Cirstea et~al.(2022)Cirstea, Guo, Yang, Kieu, Dong, and
  Pan}]{cirstea2022triformer}
Cirstea, R.-G.; Guo, C.; Yang, B.; Kieu, T.; Dong, X.; and Pan, S. 2022.
\newblock Triformer: Triangular, Variable-Specific Attentions for Long Sequence
  Multivariate Time Series Forecasting--Full Version.
\newblock \emph{arXiv preprint arXiv:2204.13767}.

\bibitem[{Derbentsev et~al.(2019)Derbentsev, Datsenko, Stepanenko, and
  Bezkorovainyi}]{derbentsev2019forecasting}
Derbentsev, V.; Datsenko, N.; Stepanenko, O.; and Bezkorovainyi, V. 2019.
\newblock Forecasting cryptocurrency prices time series using machine learning
  approach.
\newblock In \emph{SHS Web of Conferences}, volume~65, 02001. EDP Sciences.

\bibitem[{Diederik(2014)}]{diederik2014adam}
Diederik, P.~K. 2014.
\newblock Adam: A method for stochastic optimization.
\newblock \emph{(No Title)}.

\bibitem[{Garza and Mergenthaler-Canseco(2023)}]{garza2023timegpt}
Garza, A.; and Mergenthaler-Canseco, M. 2023.
\newblock TimeGPT-1.
\newblock \emph{arXiv preprint arXiv:2310.03589}.

\bibitem[{Gebru et~al.(2021)Gebru, Morgenstern, Vecchione, Vaughan, Wallach,
  Iii, and Crawford}]{gebru2021datasheets}
Gebru, T.; Morgenstern, J.; Vecchione, B.; Vaughan, J.~W.; Wallach, H.; Iii,
  H.~D.; and Crawford, K. 2021.
\newblock Datasheets for datasets.
\newblock \emph{Communications of the ACM}, 64(12): 86--92.

\bibitem[{Hamayel and Owda(2021)}]{hamayel2021novel}
Hamayel, M.~J.; and Owda, A.~Y. 2021.
\newblock A novel cryptocurrency price prediction model using GRU, LSTM and
  bi-LSTM machine learning algorithms.
\newblock \emph{Ai}, 2(4): 477--496.

\bibitem[{Han et~al.(2024)Han, Wei, Wang, Collibus, and Tessone}]{han2024mt}
Han, B.; Wei, Y.; Wang, Q.; Collibus, F. M.~D.; and Tessone, C.~J. 2024.
\newblock MT 2 AD: multi-layer temporal transaction anomaly detection in
  ethereum networks with GNN.
\newblock \emph{Complex \& Intelligent Systems}, 10(1): 613--626.

\bibitem[{Haritha and Sahana(2023)}]{haritha2023cryptocurrency}
Haritha, G.; and Sahana, N. 2023.
\newblock Cryptocurrency price prediction using twitter sentiment analysis.
\newblock In \emph{CS \& IT conference proceedings}, volume~13. CS \& IT
  Conference Proceedings.

\bibitem[{Heise et~al.(2019)Heise, Hongladarom, Jobin, {Kinder-Kurlanda}, Sun,
  Lim, Markham, Reilly, Tiidenberg, and
  Wilhelm}]{heiseInternetResearchEthical2019}
Heise, A. H.~H.; Hongladarom, S.; Jobin, A.; {Kinder-Kurlanda}, K.; Sun, S.;
  Lim, E.~L.; Markham, A.; Reilly, P.~J.; Tiidenberg, K.; and Wilhelm, C. 2019.
\newblock Internet Research: {{Ethical}} Guidelines 3.0.

\bibitem[{Hewage et~al.(2020)Hewage, Behera, Trovati, Pereira, Ghahremani,
  Palmieri, and Liu}]{hewage2020temporal}
Hewage, P.; Behera, A.; Trovati, M.; Pereira, E.; Ghahremani, M.; Palmieri, F.;
  and Liu, Y. 2020.
\newblock Temporal convolutional neural (TCN) network for an effective weather
  forecasting using time-series data from the local weather station.
\newblock \emph{Soft Computing}, 24: 16453--16482.

\bibitem[{Huang et~al.(2024)Huang, Poursafaei, Danovitch, Fey, Hu, Rossi,
  Leskovec, Bronstein, Rabusseau, and Rabbany}]{huang2024temporal}
Huang, S.; Poursafaei, F.; Danovitch, J.; Fey, M.; Hu, W.; Rossi, E.; Leskovec,
  J.; Bronstein, M.; Rabusseau, G.; and Rabbany, R. 2024.
\newblock Temporal graph benchmark for machine learning on temporal graphs.
\newblock \emph{Advances in Neural Information Processing Systems}, 36.

\bibitem[{Hyun, Lee, and Suh(2023)}]{hyun2023anti}
Hyun, W.; Lee, J.; and Suh, B. 2023.
\newblock Anti-money laundering in cryptocurrency via multi-relational graph
  neural network.
\newblock In \emph{Pacific-Asia Conference on Knowledge Discovery and Data
  Mining}, 118--130. Springer.

\bibitem[{Khan(2022)}]{khan2022graph}
Khan, A. 2022.
\newblock Graph analysis of the ethereum blockchain data: A survey of datasets,
  methods, and future work.
\newblock In \emph{2022 IEEE International Conference on Blockchain
  (Blockchain)}, 250--257. IEEE.

\bibitem[{Khedr et~al.(2021)Khedr, Arif, El-Bannany, Alhashmi, and
  Sreedharan}]{khedr2021cryptocurrency}
Khedr, A.~M.; Arif, I.; El-Bannany, M.; Alhashmi, S.~M.; and Sreedharan, M.
  2021.
\newblock Cryptocurrency price prediction using traditional statistical and
  machine-learning techniques: A survey.
\newblock \emph{Intelligent Systems in Accounting, Finance and Management},
  28(1): 3--34.

\bibitem[{Kim et~al.(2022{\natexlab{a}})Kim, Shin, Choi, and Lim}]{kim2022deep}
Kim, G.; Shin, D.-H.; Choi, J.~G.; and Lim, S. 2022{\natexlab{a}}.
\newblock A deep learning-based cryptocurrency price prediction model that uses
  on-chain data.
\newblock \emph{IEEE Access}, 10: 56232--56248.

\bibitem[{Kim et~al.(2022{\natexlab{b}})Kim, Lee, Kim, and Cho}]{kim2022graph}
Kim, J.; Lee, S.; Kim, Y.; and Cho, S. 2022{\natexlab{b}}.
\newblock A graph embedding-based identity inference attack on blockchain
  systems.
\newblock In \emph{2022 International Conference on Electronics, Information,
  and Communication (ICEIC)}, 1--3. IEEE.

\bibitem[{Kumar, Zhang, and Leskovec(2019)}]{kumar2019predicting}
Kumar, S.; Zhang, X.; and Leskovec, J. 2019.
\newblock Predicting dynamic embedding trajectory in temporal interaction
  networks.
\newblock In \emph{Proceedings of the 25th ACM SIGKDD international conference
  on knowledge discovery \& data mining}, 1269--1278.

\bibitem[{Liu et~al.(2020)Liu, Tang, Li, Guo, Fan, and Zhang}]{liu2020graph}
Liu, X.; Tang, Z.; Li, P.; Guo, S.; Fan, X.; and Zhang, J. 2020.
\newblock A graph learning based approach for identity inference in dapp
  platform blockchain.
\newblock \emph{IEEE Transactions on Emerging Topics in Computing}, 10(1):
  438--449.

\bibitem[{Liu et~al.(2023)Liu, Hu, Zhang, Wu, Wang, Ma, and
  Long}]{liu2023itransformer}
Liu, Y.; Hu, T.; Zhang, H.; Wu, H.; Wang, S.; Ma, L.; and Long, M. 2023.
\newblock itransformer: Inverted transformers are effective for time series
  forecasting.
\newblock \emph{arXiv preprint arXiv:2310.06625}.

\bibitem[{Liu et~al.(2022)Liu, Wu, Wang, and Long}]{liu2022non}
Liu, Y.; Wu, H.; Wang, J.; and Long, M. 2022.
\newblock Non-stationary transformers: Exploring the stationarity in time
  series forecasting.
\newblock \emph{Advances in Neural Information Processing Systems}, 35:
  9881--9893.

\bibitem[{Medvedev(2018)}]{medvedev2018d5}
Medvedev, E. 2018.
\newblock the D5 team.
\newblock \emph{Ethereum ETL. Available online: https://github.
  com/blockchain-etl/ethereum-etl (accessed on 25 February 2021)}.

\bibitem[{Mohamad~Zamani, Kamaruddin, and Yusof(2024)}]{mohamad2024dataset}
Mohamad~Zamani, N.~A.; Kamaruddin, N.; and Yusof, A. M.~B. 2024.
\newblock Dataset on sentiment-based cryptocurrency-related news and tweets in
  English and Malay language.
\newblock \emph{Language Resources and Evaluation}, 1--36.

\bibitem[{Nie et~al.(2022)Nie, Nguyen, Sinthong, and Kalagnanam}]{nie2022time}
Nie, Y.; Nguyen, N.~H.; Sinthong, P.; and Kalagnanam, J. 2022.
\newblock A time series is worth 64 words: Long-term forecasting with
  transformers.
\newblock \emph{arXiv preprint arXiv:2211.14730}.

\bibitem[{Ozturan, Sen, and Kilic(2021)}]{ozturan2021transaction}
Ozturan, C.; Sen, A.; and Kilic, B. 2021.
\newblock Transaction Graph Dataset for the Ethereum Blockchain.
\newblock \emph{Zenodo}, 4718440.

\bibitem[{Paszke et~al.(2019)Paszke, Gross, Massa, Lerer, Bradbury, Chanan,
  Killeen, Lin, Gimelshein, Antiga et~al.}]{paszke2019pytorch}
Paszke, A.; Gross, S.; Massa, F.; Lerer, A.; Bradbury, J.; Chanan, G.; Killeen,
  T.; Lin, Z.; Gimelshein, N.; Antiga, L.; et~al. 2019.
\newblock Pytorch: An imperative style, high-performance deep learning library.
\newblock \emph{Advances in neural information processing systems}, 32.

\bibitem[{Patel, Pan, and Rajasegarar(2020)}]{patel2020graph}
Patel, V.; Pan, L.; and Rajasegarar, S. 2020.
\newblock Graph deep learning based anomaly detection in ethereum blockchain
  network.
\newblock In \emph{International conference on network and system security},
  132--148. Springer.

\bibitem[{Poursafaei et~al.(2022)Poursafaei, Huang, Pelrine, and
  Rabbany}]{poursafaei2022towards}
Poursafaei, F.; Huang, S.; Pelrine, K.; and Rabbany, R. 2022.
\newblock Towards better evaluation for dynamic link prediction.
\newblock \emph{Advances in Neural Information Processing Systems}, 35:
  32928--32941.

\bibitem[{Qi et~al.(2023)Qi, Wu, Xu, and Guizani}]{qi2023blockchain}
Qi, Y.; Wu, J.; Xu, H.; and Guizani, M. 2023.
\newblock Blockchain Data Mining With Graph Learning: A Survey.
\newblock \emph{IEEE Transactions on Pattern Analysis and Machine
  Intelligence}.

\bibitem[{Qiu et~al.(2024)Qiu, Hu, Zhou, Wu, Du, Zhang, Guo, Zhou, Jensen,
  Sheng et~al.}]{qiu2024tfb}
Qiu, X.; Hu, J.; Zhou, L.; Wu, X.; Du, J.; Zhang, B.; Guo, C.; Zhou, A.;
  Jensen, C.~S.; Sheng, Z.; et~al. 2024.
\newblock Tfb: Towards comprehensive and fair benchmarking of time series
  forecasting methods.
\newblock \emph{arXiv preprint arXiv:2403.20150}.

\bibitem[{Rossi et~al.(2020)Rossi, Chamberlain, Frasca, Eynard, Monti, and
  Bronstein}]{rossi2020temporal}
Rossi, E.; Chamberlain, B.; Frasca, F.; Eynard, D.; Monti, F.; and Bronstein,
  M. 2020.
\newblock Temporal graph networks for deep learning on dynamic graphs.
\newblock \emph{arXiv preprint arXiv:2006.10637}.

\bibitem[{Sepehri et~al.(2025)Sepehri, Mehradfar, Soltanolkotabi, and
  Avestimehr}]{sepehri2025cryptomamba}
Sepehri, M.~S.; Mehradfar, A.; Soltanolkotabi, M.; and Avestimehr, S. 2025.
\newblock CryptoMamba: Leveraging State Space Models for Accurate Bitcoin Price
  Prediction.
\newblock \emph{arXiv preprint arXiv:2501.01010}.

\bibitem[{Shamsi et~al.(2022)Shamsi, Victor, Kantarcioglu, Gel, and
  Akcora}]{shamsi2022chartalist}
Shamsi, K.; Victor, F.; Kantarcioglu, M.; Gel, Y.; and Akcora, C.~G. 2022.
\newblock Chartalist: Labeled graph datasets for utxo and account-based
  blockchains.
\newblock \emph{Advances in Neural Information Processing Systems}, 35:
  34926--34939.

\bibitem[{Sherstinsky(2020)}]{sherstinsky2020fundamentals}
Sherstinsky, A. 2020.
\newblock Fundamentals of recurrent neural network (RNN) and long short-term
  memory (LSTM) network.
\newblock \emph{Physica D: Nonlinear Phenomena}, 404: 132306.

\bibitem[{Shumway et~al.(2017)Shumway, Stoffer, Shumway, and
  Stoffer}]{shumway2017arima}
Shumway, R.~H.; Stoffer, D.~S.; Shumway, R.~H.; and Stoffer, D.~S. 2017.
\newblock ARIMA models.
\newblock \emph{Time series analysis and its applications: with R examples},
  75--163.

\bibitem[{Townsend and Wallace(2016)}]{townsendSocialMediaResearch2016}
Townsend, L.; and Wallace, C. 2016.
\newblock Social Media Research: {{A}} Guide to Ethics.
\newblock \emph{University of Aberdeen}, 1(16): 1--16.

\bibitem[{Trivedi et~al.(2019)Trivedi, Farajtabar, Biswal, and
  Zha}]{trivedi2019dyrep}
Trivedi, R.; Farajtabar, M.; Biswal, P.; and Zha, H. 2019.
\newblock Dyrep: Learning representations over dynamic graphs.
\newblock In \emph{International conference on learning representations}.

\bibitem[{Uhlig(2022)}]{uhlig2022luna}
Uhlig, H. 2022.
\newblock A luna-tic stablecoin crash.
\newblock Technical report, National Bureau of Economic Research.

\bibitem[{Wang et~al.(2023)Wang, Peng, Huang, Wang, Chen, and
  Xiao}]{wang2023micn}
Wang, H.; Peng, J.; Huang, F.; Wang, J.; Chen, J.; and Xiao, Y. 2023.
\newblock Micn: Multi-scale local and global context modeling for long-term
  series forecasting.
\newblock In \emph{The eleventh international conference on learning
  representations}.

\bibitem[{Wang et~al.(2021{\natexlab{a}})Wang, Chang, Li, Chu, Li, Zhang, He,
  Song, Zhou, and Yang}]{wang2021tcl}
Wang, L.; Chang, X.; Li, S.; Chu, Y.; Li, H.; Zhang, W.; He, X.; Song, L.;
  Zhou, J.; and Yang, H. 2021{\natexlab{a}}.
\newblock Tcl: Transformer-based dynamic graph modelling via contrastive
  learning.
\newblock \emph{arXiv preprint arXiv:2105.07944}.

\bibitem[{Wang et~al.(2024{\natexlab{a}})Wang, Zhang, Liu, Lu, Luo, and
  He}]{wangex}
Wang, Q.; Zhang, Z.; Liu, Z.; Lu, S.; Luo, B.; and He, B. 2024{\natexlab{a}}.
\newblock EX-Graph: A Pioneering Dataset Bridging Ethereum and X.
\newblock In \emph{The Twelfth International Conference on Learning
  Representations}.

\bibitem[{Wang et~al.(2024{\natexlab{b}})Wang, Wu, Shi, Hu, Luo, Ma, Zhang, and
  Zhou}]{wang2024timemixer}
Wang, S.; Wu, H.; Shi, X.; Hu, T.; Luo, H.; Ma, L.; Zhang, J.~Y.; and Zhou, J.
  2024{\natexlab{b}}.
\newblock Timemixer: Decomposable multiscale mixing for time series
  forecasting.
\newblock \emph{arXiv preprint arXiv:2405.14616}.

\bibitem[{Wang et~al.(2021{\natexlab{b}})Wang, Chang, Liu, Leskovec, and
  Li}]{wang2021inductive}
Wang, Y.; Chang, Y.-Y.; Liu, Y.; Leskovec, J.; and Li, P. 2021{\natexlab{b}}.
\newblock Inductive representation learning in temporal networks via causal
  anonymous walks.
\newblock \emph{arXiv preprint arXiv:2101.05974}.

\bibitem[{Wirawan, Widiyaningtyas, and Hasan(2019)}]{wirawan2019short}
Wirawan, I.~M.; Widiyaningtyas, T.; and Hasan, M.~M. 2019.
\newblock Short term prediction on bitcoin price using ARIMA method.
\newblock In \emph{2019 International Seminar on Application for Technology of
  Information and Communication (iSemantic)}, 260--265. IEEE.

\bibitem[{Wu et~al.(2022)Wu, Hu, Liu, Zhou, Wang, and Long}]{wu2022timesnet}
Wu, H.; Hu, T.; Liu, Y.; Zhou, H.; Wang, J.; and Long, M. 2022.
\newblock Timesnet: Temporal 2d-variation modeling for general time series
  analysis.
\newblock \emph{arXiv preprint arXiv:2210.02186}.

\bibitem[{Xu et~al.(2020)Xu, Ruan, Korpeoglu, Kumar, and
  Achan}]{xu2020inductive}
Xu, D.; Ruan, C.; Korpeoglu, E.; Kumar, S.; and Achan, K. 2020.
\newblock Inductive representation learning on temporal graphs.
\newblock \emph{arXiv preprint arXiv:2002.07962}.

\bibitem[{Xu, Zeng, and Xu(2023)}]{xu2023fits}
Xu, Z.; Zeng, A.; and Xu, Q. 2023.
\newblock FITS: Modeling time series with $10 k $ parameters.
\newblock \emph{arXiv preprint arXiv:2307.03756}.

\bibitem[{Yenido{\u{g}}an et~al.(2018)Yenido{\u{g}}an, {\c{C}}ayir, Kozan,
  Da{\u{g}}, and Arslan}]{yenidougan2018bitcoin}
Yenido{\u{g}}an, I.; {\c{C}}ayir, A.; Kozan, O.; Da{\u{g}}, T.; and Arslan,
  {\c{C}}. 2018.
\newblock Bitcoin forecasting using ARIMA and PROPHET.
\newblock In \emph{2018 3rd international conference on computer science and
  engineering (UBMK)}, 621--624. IEEE.

\bibitem[{Yu et~al.(2023)Yu, Sun, Du, and Lv}]{yu2023towards}
Yu, L.; Sun, L.; Du, B.; and Lv, W. 2023.
\newblock Towards better dynamic graph learning: New architecture and unified
  library.
\newblock \emph{Advances in Neural Information Processing Systems}, 36:
  67686--67700.

\bibitem[{Zeng et~al.(2023)Zeng, Chen, Zhang, and Xu}]{zeng2023transformers}
Zeng, A.; Chen, M.; Zhang, L.; and Xu, Q. 2023.
\newblock Are transformers effective for time series forecasting?
\newblock In \emph{Proceedings of the AAAI conference on artificial
  intelligence}, volume~37, 11121--11128.

\bibitem[{Zhang and Yan(2023)}]{zhang2023crossformer}
Zhang, Y.; and Yan, J. 2023.
\newblock Crossformer: Transformer utilizing cross-dimension dependency for
  multivariate time series forecasting.
\newblock In \emph{The eleventh international conference on learning
  representations}.

\bibitem[{Zhang et~al.(2023)Zhang, Luo, Lu, and He}]{zhang2023live}
Zhang, Z.; Luo, B.; Lu, S.; and He, B. 2023.
\newblock Live graph lab: Towards open, dynamic and real transaction graphs
  with NFT.
\newblock \emph{Advances in Neural Information Processing Systems}, 36:
  18769--18793.

\bibitem[{Zhengyang et~al.(2019)Zhengyang, Xingzhou, Jinjin, and
  Jiaqing}]{zhengyang2019prediction}
Zhengyang, W.; Xingzhou, L.; Jinjin, R.; and Jiaqing, K. 2019.
\newblock Prediction of cryptocurrency price dynamics with multiple machine
  learning techniques.
\newblock In \emph{Proceedings of the 2019 4th International Conference on
  Machine Learning Technologies}, 15--19.

\bibitem[{Zhou et~al.(2021)Zhou, Zhang, Peng, Zhang, Li, Xiong, and
  Zhang}]{zhou2021informer}
Zhou, H.; Zhang, S.; Peng, J.; Zhang, S.; Li, J.; Xiong, H.; and Zhang, W.
  2021.
\newblock Informer: Beyond efficient transformer for long sequence time-series
  forecasting.
\newblock In \emph{Proceedings of the AAAI conference on artificial
  intelligence}, volume~35, 11106--11115.

\bibitem[{Zhou et~al.(2022{\natexlab{a}})Zhou, Ma, Wen, Sun, Yao, Yin, Jin
  et~al.}]{zhou2022film}
Zhou, T.; Ma, Z.; Wen, Q.; Sun, L.; Yao, T.; Yin, W.; Jin, R.; et~al.
  2022{\natexlab{a}}.
\newblock Film: Frequency improved legendre memory model for long-term time
  series forecasting.
\newblock \emph{Advances in neural information processing systems}, 35:
  12677--12690.

\bibitem[{Zhou et~al.(2022{\natexlab{b}})Zhou, Ma, Wen, Wang, Sun, and
  Jin}]{zhou2022fedformer}
Zhou, T.; Ma, Z.; Wen, Q.; Wang, X.; Sun, L.; and Jin, R. 2022{\natexlab{b}}.
\newblock Fedformer: Frequency enhanced decomposed transformer for long-term
  series forecasting.
\newblock In \emph{International conference on machine learning}, 27268--27286.
  PMLR.

\bibitem[{Zou, Tuncali, and Silverman(2003)}]{zou2003correlation}
Zou, K.~H.; Tuncali, K.; and Silverman, S.~G. 2003.
\newblock Correlation and simple linear regression.
\newblock \emph{Radiology}, 227(3): 617--628.

\bibitem[{Zoumpekas, Houstis, and Vavalis(2020)}]{zoumpekas2020eth}
Zoumpekas, T.; Houstis, E.; and Vavalis, M. 2020.
\newblock ETH analysis and predictions utilizing deep learning.
\newblock \emph{Expert Systems with Applications}, 162: 113866.

\end{thebibliography}

\bigskip 

\newpage

\section{Ethics Checklist}

\begin{enumerate}

\item For most authors...
\begin{enumerate}
    \item  Would answering this research question advance science without violating social contracts, such as violating privacy norms, perpetuating unfair profiling, exacerbating the socio-economic divide, or implying disrespect to societies or cultures?
    \answerYes{Yes} 
  \item Do your main claims in the abstract and introduction accurately reflect the paper's contributions and scope?
    \answerYes{Yes} 
   \item Do you clarify how the proposed methodological approach is appropriate for the claims made? 
    \answerYes{Yes} 
   \item Do you clarify what are possible artifacts in the data used, given population-specific distributions?
    \answerYes{Yes} 
  \item Did you describe the limitations of your work?
    \answerYes{Yes} 
  \item Did you discuss any potential negative societal impacts of your work?
    \answerYes{Yes} 
      \item Did you discuss any potential misuse of your work?
    \answerYes{Yes} 
    \item Did you describe steps taken to prevent or mitigate potential negative outcomes of the research, such as data and model documentation, data anonymization, responsible release, access control, and the reproducibility of findings?
    \answerYes{Yes} 
  \item Have you read the ethics review guidelines and ensured that your paper conforms to them?
    \answerYes{Yes} 
\end{enumerate}

\item Additionally, if your study involves hypotheses testing...
\begin{enumerate}
  \item Did you clearly state the assumptions underlying all theoretical results?
    \answerNA{NA}
  \item Have you provided justifications for all theoretical results?
    \answerNA{NA}
  \item Did you discuss competing hypotheses or theories that might challenge or complement your theoretical results?
    \answerNA{NA}
  \item Have you considered alternative mechanisms or explanations that might account for the same outcomes observed in your study?
    \answerNA{NA}
  \item Did you address potential biases or limitations in your theoretical framework?
    \answerNA{NA}
  \item Have you related your theoretical results to the existing literature in social science?
    \answerNA{NA}
  \item Did you discuss the implications of your theoretical results for policy, practice, or further research in the social science domain?
    \answerNA{NA}
\end{enumerate}

\item Additionally, if you are including theoretical proofs...
\begin{enumerate}
  \item Did you state the full set of assumptions of all theoretical results?
    \answerNA{NA}
	\item Did you include complete proofs of all theoretical results?
    \answerNA{NA}
\end{enumerate}

\item Additionally, if you ran machine learning experiments...
\begin{enumerate}
  \item Did you include the code, data, and instructions needed to reproduce the main experimental results (either in the supplemental material or as a URL)?
    \answerYes{Yes} 
  \item Did you specify all the training details (e.g., data splits, hyperparameters, how they were chosen)?
    \answerYes{Yes} 
     \item Did you report error bars (e.g., with respect to the random seed after running experiments multiple times)?
    \answerYes{Yes} 
	\item Did you include the total amount of compute and the type of resources used (e.g., type of GPUs, internal cluster, or cloud provider)?
    \answerYes{Yes} 
     \item Do you justify how the proposed evaluation is sufficient and appropriate to the claims made? 
    \answerYes{Yes} 
     \item Do you discuss what is ``the cost`` of misclassification and fault (in)tolerance?
    \answerYes{Yes} 
  
\end{enumerate}

\item Additionally, if you are using existing assets (e.g., code, data, models) or curating/releasing new assets, \textbf{without compromising anonymity}...
\begin{enumerate}
  \item If your work uses existing assets, did you cite the creators?
    \answerYes{Yes} 
  \item Did you mention the license of the assets?
    \answerYes{Yes} 
  \item Did you include any new assets in the supplemental material or as a URL?
    \answerYes{Yes} 
  \item Did you discuss whether and how consent was obtained from people whose data you're using/curating?
    \answerYes{Yes} 
  \item Did you discuss whether the data you are using/curating contains personally identifiable information or offensive content?
    \answerYes{Yes} 
\item If you are curating or releasing new datasets, did you discuss how you intend to make your datasets FAIR (see \citet{fair})?
\answerYes{Yes} 
\item If you are curating or releasing new datasets, did you create a Datasheet for the Dataset (see \citet{gebru2021datasheets})? 
\answerYes{Yes} 
\end{enumerate}

\item Additionally, if you used crowdsourcing or conducted research with human subjects, \textbf{without compromising anonymity}...
\begin{enumerate}
  \item Did you include the full text of instructions given to participants and screenshots?
    \answerNA{NA}
  \item Did you describe any potential participant risks, with mentions of Institutional Review Board (IRB) approvals?
    \answerNA{NA}
  \item Did you include the estimated hourly wage paid to participants and the total amount spent on participant compensation?
    \answerNA{NA}
   \item Did you discuss how data is stored, shared, and deidentified?
   \answerNA{NA}
\end{enumerate}

\end{enumerate}

\newpage

\text{\qquad }

\appendix
\noindent {\bf \Large Appendix}

\section{Datasheets for Datasets}

\subsection*{1. Motivation}

\paragraph{What is the purpose of creating the dataset? Is it designed for specific tasks?}  
The dataset was created to support in-depth analysis of the cryptocurrency market, particularly for tasks such as user behavior prediction and token price prediction. By integrating multi-source data, including transaction records, token information, market indices, and sentiment data, it provides a platform for studying user behavior, market fluctuations, and associated risks.

\paragraph{Who created the dataset (e.g., team, institution)?}  
The dataset was created by the Liu AI Lab team at the University of Auckland, under the platform name 3MEthTaskforce.

\subsection*{2. Composition}

\paragraph{What do the dataset instances represent (e.g., documents, images)?}  
The dataset instances represent blockchain transaction records, token information, market indices, and textual sentiment data from Reddit.

\paragraph{How many instances are there in total?}  
The dataset contains 303 million transaction records, covering 35 million users and 3,880 tokens.

\paragraph{Is the dataset a complete set or a sample from a larger set?}  
The dataset is a multi-source collection based on the Ethereum network and is not a sample from a larger set.

\paragraph{What data does each instance contain?}  
Each instance includes information such as sender and receiver wallet addresses, token addresses, transaction value, and blockchain timestamps.

\paragraph{Does any instance lack information?}  
The paper does not explicitly mention missing information in the instances, but it states that the raw data underwent cleaning and annotation to ensure completeness.

\paragraph{Are there recommended data splits (e.g., training, validation, testing)?}  
The dataset is recommended to be split into training (70\%), validation (15\%), and testing (15\%) sets.

\paragraph{Are there errors, noise, or redundancies in the dataset?}  
The dataset is free from errors, noise, and redundancies.

\subsection*{3. Collection Process}

\paragraph{How was the data for each instance acquired (e.g., direct observation, subject reports)?}  
The data was collected using various open-source tools and platforms, including the Ethereum Public ETL tool, DefiLlama, and Reddit API (PRAW).

\paragraph{What mechanisms or procedures were used for data collection?}  
The data was primarily obtained using blockchain extraction tools, transaction APIs, and sentiment analysis APIs.

\paragraph{What is the time range of data collection?}  
The dataset covers transaction records and sentiment data from 2014 to 2024.

\paragraph{Were any ethical review processes conducted?}  
The data collection followed ethical guidelines for the use of publicly available data, requiring no additional ethical review.

\subsection*{4. Preprocessing/Cleaning/Annotation}

\paragraph{Was any preprocessing, cleaning, or annotation performed on the data?}  
Yes, the data was cleaned and annotated, including the removal of irrelevant content and the extraction of sentiment indices.

\paragraph{Was the “raw” data retained?}  
Yes, the raw data was retained for subsequent verification and further analysis.

\subsection*{5. Use Cases}

\paragraph{For which tasks has the dataset been used?}  
The dataset has been used for tasks such as user behavior prediction, token price prediction, and user behavior marking.

\paragraph{How might the dataset’s composition or collection process affect its future uses?}  
The multi-source and multi-modal nature of the dataset makes it suitable for a broader range of cryptocurrency analysis tasks, such as risk forecasting and market behavior analysis.

\paragraph{Are there any tasks for which the dataset should not be used?}  
The dataset is not suitable for tasks unrelated to blockchain or for data analysis scenarios requiring strict privacy protection.

\subsection*{6. Distribution}

\paragraph{Will the dataset be distributed to third parties?}  
Yes, the dataset has been made publicly available.

\paragraph{How will the dataset be distributed (e.g., tarball on a website, API)?}  
The dataset is provided in CSV format through the Figshare platform, with a permanent DOI link.

\paragraph{Are there any copyright or usage terms?}  
The dataset is distributed under the CC BY 4.0 license, requiring proper attribution for use.

\subsection*{7. Maintenance}

\paragraph{Who is responsible for maintaining the dataset?}  
The dataset is maintained by the creation team.

\paragraph{Will the dataset be updated?}  
Yes, updates and extensions may be made in the future as needed.

\paragraph{Is there a mechanism for others to contribute to or extend the dataset?}  
Contributions or extensions can be made under the guidance of the original team.

\section{Ethics and Privacy}
\label{app:ethics}

This research adheres to the ethical guidelines outlined by the Association of Internet Researchers  \cite{heiseInternetResearchEthical2019} and Townsend \& Wallace  \cite{townsendSocialMediaResearch2016}. In conducting this study, we have carefully considered several factors to minimize potential ethical and privacy risks.

Ethereum is a public and permissionless blockchain, which means its entire network is open for anyone to join, participate in consensus, execute transactions, and view the ledger. This openness underscores its public nature, with no central authority controlling access to the network or its data. Similarly, the Reddit data used in this study is sourced exclusively from public subreddits and posts. In line with the emphasis on publicness, no private messages or content from restricted communities have been accessed or included.

In maintaining the pseudonymity of transactions, our dataset preserves the same level of anonymity inherent to the Ethereum network. Reddit users typically operate under pseudonyms, and to further minimize privacy risks, we have anonymized the data by replacing post identifiers with randomly generated values. In addition, we applied data processing techniques to mask any personally identifiable information (PII) and sensitive information that might have been present in the raw data.

We have implemented strict protocols governing the use and distribution of this dataset. Researchers seeking access will be required to agree to terms that prohibit any attempts to re-identify individuals or use the data for purposes other than approved research.

The insights gained from this dataset have the potential to contribute significantly to our understanding of cryptocurrency trading and its broader ecosystems. These findings can help advance the development of more secure, equitable, and robust blockchain and decentralized finance (DeFi) systems. We believe the potential benefits of this research outweigh the minimal risks associated with using publicly available data.

\section{Raw Data}
\label{row_data}




Table \ref{tab:token global total market cap} is an example of the Market Cap in the Global Index Section. It shows the time-series data of the total market capitalization of cryptocurrencies, covering five dates from April 29, 2013, to May 7, 2013.

\begin{table}[H]

\centering
\caption{Examples of Total Market Capitalization Data}\label{tab:token global total market cap}
\begin{tabular}{lc}
\toprule
\textbf{DateTime} & \textbf{Market Cap} \\ 
\midrule
2013/4/29 12:00  & 1583440000 \\ 
2013/5/1 12:00   & 1637389952 \\ 
2013/5/3 12:00   & 1275410048 \\ 
2013/5/5 12:00   & 1335379968 \\ 
2013/5/7 12:00   & 1313900032 \\ 
\bottomrule
\end{tabular}
\label{tab:market_cap_data}
\end{table}

Table \ref{tab:token global 24th volume} is an example of the 24h Volume Data in the Global Index Section. This table presents the 24-hour trading volume from February 25, 2014, to March 5, 2014. The volume ranges from 7,047,3200 to 11,784,0000, indicating fluctuations in market activity during this period.

\begin{table}[H]

\centering
\caption{Examples of 24h Volume Data}\label{tab:token global 24th volume}
\begin{tabular}{lc}
\toprule
\textbf{DateTime} & \textbf{Volume (24h)} \\ 
\midrule
2014/2/25 13:00  & 70473200  \\ 
2014/2/27 13:00  & 84957000  \\ 
2014/3/1 13:00   & 41190400  \\ 
2014/3/3 13:00   & 17991000  \\ 
2014/3/5 13:00   & 117840000 \\ 
\bottomrule
\end{tabular}
\label{tab:24h_volume_data}
\end{table}

Table \ref{tab:token global stable coin} is an example of the Stablecoin Market Capitalization Data in the Global Index Section. It shows the stablecoin market capitalization data from March 10, 2016, to March 18, 2016.

\begin{table}[H]

\centering
\caption{Examples of Stablecoin Market Capitalization Data}\label{tab:token global stable coin}
\begin{tabular}{lc}
\toprule
\textbf{DateTime} & \textbf{Stablecoin Market Cap} \\ 
\midrule
2016/3/10 13:00  & 1451448.067 \\ 
2016/3/12 13:00  & 1451593.424 \\ 
2016/3/14 13:00  & 1451479.734 \\ 
2016/3/16 13:00  & 1451602.250 \\ 
2016/3/18 13:00  & 1451575.947 \\ 
\bottomrule
\end{tabular}
\label{tab:stablecoin_market_cap}
\end{table}




Table \ref{tab:token transaction} is an example of Token Transaction Data for token AAVE. This table records the transaction information for AAVE, including token address, sender and receiver addresses, transaction value, transaction hash, log index, block timestamp, and block number. The transaction times range from the early hours of July 26, 2024, to later that night, showcasing multiple large transactions of AAVE tokens.

\begin{table*}[h]
\centering
\caption{Examples in Token Transaction Data for AAVE}\label{tab:token transaction}
\begin{tabular}{lccccccc}
\toprule
\textbf{TokenAdd} & \textbf{FromAdd} & \textbf{ToAdd} & \textbf{Value} & \textbf{TransactionHash} & \textbf{Index} & \textbf{BlockTimestamp} & \textbf{BlockNumber} \\ 
\midrule
0x...dae9 & 0x...7fad & 0x...7e1c & 8.40279E+16 & 0x...276e3 & 101 & 2024-07-26 01:28:59 & 20387439 \\ 
0x...dae9 & 0x...5622 & 0x...9a81 & 8.51435E+18 & 0x...90c4c & 484 & 2024-07-26 01:14:11 & 20387365 \\ 
0x...dae9 & 0x...f2c8 & 0x...4bee & 1.15726E+20 & 0x...13aac & 171 & 2024-07-26 01:12:11 & 20387356 \\ 
0x...dae9 & 0x...5145 & 0x...0703 & 1.14393E+21 & 0x...ac8cf & 367 & 2024-07-25 20:47:23 & 20386044 \\ 
0x...dae9 & 0x...699c & 0x...5fb6 & 3.40423E+18 & 0x...84f0 & 204 & 2024-07-25 20:10:47 & 20385863 \\ 
\bottomrule
\end{tabular}
\label{tab:masked_token_transactions}
\end{table*}

Table \ref{tab:token info token general} is an example of the token\_general Table in the Token Info Section. This table lists information about five tokens, including ID, symbol, name, Ethereum address, and decimals.

\begin{table*}[h]

\centering
\caption{Examples in token\_general\_3880.csv Table}\label{tab:token info token general}
\begin{tabular}{lcccc}
\toprule
\textbf{ID} & \textbf{Symbol} & \textbf{Name} & \textbf{ETH Address} & \textbf{Decimal} \\ 
\midrule
0 & zcn  & Zus                       & ...38f3b78 & 10  \\ 
1 & 0kn  & 0 Knowledge Network        & ...7d29036 & 18  \\ 
2 & ome  & O-MEE                     & ...826977e & 18  \\ 
3 & zrx  & 0x Protocol                & ...699f498 & 18  \\ 
4 & 0x0  & 0x0.ai: AI Smart Contract  & ...0811ad5 & 9   \\ 
\bottomrule
\end{tabular}
\label{tab:token_info}
\end{table*}

Table \ref{tab:token info token recording} is an example of Token Recording BNB in the Token Info Section. It records the historical price, market cap, and total volume for BNB tokens. The data covers timestamps, prices, market capitalizations, and total volumes across several time intervals.

\begin{table*}[h]

\centering
\caption{Examples in Token History Information for BNB ...1bdd52.csv}\label{tab:token info token recording}
\begin{tabular}{lcccc}
\toprule
\textbf{Timestamp} & \textbf{Price} & \textbf{Market Caps} & \textbf{Total Volumes} \\ 
\midrule
1.50552E+12 & 0.107250624 & 10725062.44 & 1.051223307 \\ 
1.50561E+12 & 0.154041291 & 15404129.09 & 14.67858722 \\ 
1.50569E+12 & 0.173491239 & 17349123.91 & 6.001766938 \\ 
1.50578E+12 & 0.168334191 & 16833419.06 & 3.878927407 \\ 
1.50587E+12 & 0.166627925 & 16662792.49 & 40.6876186 \\ 
\bottomrule
\end{tabular}
\label{tab:price_marketcap_volume}
\end{table*}

Table \ref{tab:token sentiment} is an example of the Token Sentiment Data using background knowledge from an LLM for sentiment scoring. The table provides sentiment analysis of textual data, including score, timestamp, number of comments, text content, and positive/negative sentiment scores. For instance, a comment on November 19, 2013, received 3 positive and 7 negative scores, with the text partially masked.

\begin{table*}[h]

\centering
\caption{Examples in Text Data with Sentiment Analysis}\label{tab:token sentiment}
\begin{tabular}{lcccccc}
\toprule
\textbf{Score} & \textbf{Timestamp} & \textbf{Number of Comments} & \textbf{Text (masked)} & \textbf{Positive} & \textbf{Negative} \\ 
\midrule
3191 & 2013/11/19 19:15 & 472 & I’m one of the Senators attending... & 3 & 7 \\ 
3193 & 2013/11/25 1:38 & 282 & I was bored so I animated the... & 1 & 0 \\ 
3524 & 2014/2/13 23:49 & 470 & on r/bitcoin right now... & 0 & 7 \\ 
3055 & 2014/2/18 20:15 & 463 & Bitcoin takes a walk with Dogecoin... & 2 & 1 \\ 
3455 & 2014/2/26 16:41 & 489 & Open Letter to Michael Casey - WSJ reporter... & 5 & 7 \\ 
3954 & 2014/2/28 8:17 & 416 & We've gotta be able to laugh at ourselves... & 1 & 0 \\ 
\bottomrule
\end{tabular}
\label{tab:masked_text_data}
\end{table*}

Table \ref{tab:crypto_market_cap} is an example of Bitcoin Dominance Data in the Global Index Section. It includes the market capitalizations of Bitcoin (BTC), Ethereum (ETH), Tether (USDT), BNB, Solana (SOL), and other cryptocurrencies.

\begin{table*}[h]
\centering
\caption{Examples in Cryptocurrency Bitcoin Dominance Data}\label{tab:crypto_market_cap}
\begin{tabular}{lcccccc}
\toprule
\textbf{DateTime} & \textbf{BTC Cap} & \textbf{ETH Cap} & \textbf{USDT Cap} & \textbf{BNB Cap} & \textbf{SOL Cap} & \textbf{Others Cap} \\ 
\midrule
2023/9/3 12:00  & 5.04E+11  & 1.97E+11  & 82901682430   & 32993907647  & 7967342850  & 2.18E+11  \\ 
2023/9/10 12:00 & 5.04E+11  & 1.97E+11  & 82992946191   & 32957692204  & 7984825883  & 2.19E+11  \\ 
2023/9/17 12:00 & 5.18E+11  & 1.97E+11  & 83065027653   & 33069547650  & 7871412204  & 2.22E+11  \\ 
2023/9/24 13:00 & 5.18E+11  & 1.92E+11  & 83206758665   & 32396394279  & 8035147376  & 2.21E+11  \\ 
2023/10/1 13:00 & 5.26E+11  & 2.01E+11  & 83260095938   & 33046306303  & 8834373543  & 2.25E+11  \\ 
\bottomrule
\end{tabular}

\end{table*}

\section{User Behaviour Experiment Setup} We aim to compare different GNN models, input features, and the impact of two types of sentiment features on performance.

The set of vertices $V$ of the graph is $U\cup C$ containing both the set of users and tokens. A {\em temporal edge} is of the form $(u,c_j,t)$ where $u\in U$, $c_j\in C$, and $1\leq t\leq \tau$ is the timestamp of the edge. This edge denotes a transaction where user $u$ purchases some amount of token $c_j$ at timestamp $t$. 
For each temporal edge $(u,c_j,t)$, associate $q$ and $\mathrm{trans}_{j,t}$ to form a {\em labelled temporal edge} $(u,c,q,\mathrm{trans}_{j,t})$, where $q$ represents the amount of token $c_j$ purchased during this transaction, and $\mathrm{trans}_{j,t}$ is a {\em transaction label}, discussed below. 

For the user behavior prediction task, we construct a graph $(V, E)$ where $E$ contains the set of all such labelled temporal edges up to a certain timestamp $\tau$. This graph can be constructed using the dataset in 3MEthTaskforce. 
Our goal is to train a GNN that predicts, for any $u\in U$ and $c\in C$ and a future timestamp $t'>\tau$, whether a temporal edge $(u,c,t')$ will appear. 

To describe our baseline methods in detail, we need to elaborate on the following issues. 

\paragraph*{\bf Textual Sentiment Index Extraction.} To convert textual data into sentiment indices, we concatenate the top posts at time \( t \) into a vector \( {\bf txt}_t = \text{txt}_{t,1} \mid \text{txt}_{t,2} \mid \dots \mid \text{txt}_{t,\ell} \). We then apply two methods to extract sentiment indices from \( {\bf txt}_t \):

\noindent{\bf (1)} Following  \cite{bhatt2023machine}, we use a large language model (DeepSeek) to generate an overall sentiment score \( s^{\text{overall}}_t \) on a scale of 0 to 10, where 5 represents neutral sentiment, below 5 indicates negative sentiment, and above 5 indicates positive sentiment.

\noindent{\bf (2)} The second method refines the first by incorporating the post timestamps to account for cryptocurrency trends and generating two separate scores: \( s^{\text{pos}}_t \) for positive sentiment and \( s^{\text{neg}}_t \) for negative sentiment.

\paragraph*{\bf Transaction Labels.} To comprehensively analyze performance of models for this link prediction task, we define several different transaction label vectors:
\begin{itemize}
    \item $\text{trans\_record}_{j,t} = (p_{j,t}, m_{j,t}, v_{j,t})$ includes only token information. 
    \item $\text{trans\_global}_{j,t} = g_t$ includes only global market index.
    \item $\text{trans\_text}_{j,t} = s^{\text{overall}}_t$ includes the only overall sentiment score.
    \item $\text{trans\_text\_llm}_{j,t} = (s^{\text{pos}}_t, s^{\text{neg}}_t)$ contains both positive and negative sentiment scores.
    \item $\text{trans\_all}_{j,t} = (p_{j,t}, m_{j,t}, v_{j,t}, g_t, s^{\text{overall}}_t,  s^{\text{pos}}_t, s^{\text{neg}}_t)$ includes all features:  token information, global market index, and sentiment scores.
\end{itemize}

\paragraph{Dataset}
We use the smaller dataset from the Transaction Record section, containing approximately $260,000$ transactions and $29,164$ active wallet addresses. From this, we extract the transaction label vectors as defined above. The datasets are split chronologically into train/validation/test sets with a 70\%/15\%/15\% ratio.

\paragraph{Parameter Settings}
We train models using Adam \cite{diederik2014adam} with binary cross-entropy loss. All models are trained for 100 epochs, with early stopping after 20 epochs of no improvement. The learning rate is set to 0.0001, and the batch size is 200.

\paragraph{Performance Metrics}
Following \cite{xu2020inductive, poursafaei2022towards, wang2021inductive, rossi2020temporal}, we evaluate user behavior prediction using test set average precision (TAP) and new node average precision (NAP). The inductive negative sampling strategy is described in \cite{poursafaei2022towards}. All results are averaged over three runs.

\section{GNN Model}
\label{app:GNN}

\paragraph{DyGFormer} lies in introducing the self-attention mechanism of Transformers to capture long-term dependencies in dynamic graphs. By incorporating temporal encoding within the Transformer, DyGFormer effectively captures complex dynamic changes in graphs. Compared to other models that utilize short-term temporal windows, DyGFormer excels in tasks involving long-term dependencies.

\paragraph{JODIE} focuses on continuous time-series interaction modeling, tracking the long-term interaction dynamics between users and items. This bidirectional embedding update mechanism makes it particularly effective in scenarios with frequent user-item interactions (e.g., recommendation systems). JODIE leverages RNNs to capture the historical dependencies of nodes (users and items), making it well-suited for long-term behavior prediction.

\paragraph{DyRep} lies in its distinction between two types of events: interaction events between nodes (such as transactions or communications) and structural events (such as the creation or deletion of nodes/edges). DyRep divides dynamic graph tasks into these two categories of events and models them using event time sequences, making it ideal for tasks involving constantly changing nodes and edges.

\paragraph{TGAT} is the combination of Graph Convolutional Networks (GCNs) with temporal encoding and the use of graph attention mechanisms to capture dynamic relationships in the graph. Unlike simple temporal updates, TGAT focuses on important neighboring nodes through attention mechanisms, making it well-suited for dynamic networks with frequent interactions.

\paragraph{TGN} introduces a message-passing mechanism to update node states, making it particularly suitable for learning node embeddings in large-scale dynamic graphs. Unlike TGAT, TGN is not restricted to attention mechanisms but combines message-passing and temporal updates, allowing it to scale to large graphs and handle complex dynamic environments.

\paragraph{TCL} utilizes contrastive learning to capture the temporal evolution of node embeddings in dynamic graphs by comparing node representations at different time points. Unlike other models that rely on supervised learning, TCL constructs positive and negative sample pairs (e.g., the same node at different time points) to capture temporal changes. This unsupervised learning approach enables TCL to perform well in dynamic graph tasks without explicit labels.

\section{Prompt}
\label{app:prompt}

\paragraph{Prompt 1: }You are a useful cryptocurrency social media post sentiment analysis expert. Now I give you the text of the top reddit posts on cryptocurrency scope, then you need to understand and analyze the sentiment score of this post in the context of cryptocurrency (on a scale of 10, 0-4 is negative, higher is positive, lower is negative, 6-10 is positive, 5 is neutral).Below is the text of reddit's top crypto-scoped posts that you should analyze as above, just give me the final score, just give a number like "6" :

\paragraph{Prompt 2: }You are a helpful cryptocurrency social media posts sentiment analysis expert. Now I give you the cryptocurrency scope top posts' text of reddit with its timestamp, and then you need understand and analyse the following content in context of cryptocurrency with considering the trends and news about cryptocurrencies at the time of the timestamp and give me the negative score and the positive score. The sentiment score may take into account not only the sentiment of the post regarding cryptocurrencies, but also whether the time of the post and the context (e.g., big events in the cryptocurrency space) seemed to have a positive or negative impact on cryptocurrencies at the time. For example, if the text say: Bitcoin is a kind of cryptocurrency, this is a totally neutral, so the negative score and the positive score are 0; if the text say: I like Bitcoin, while bitcoin has high volatility; this somewhat negative and somewhat positive, so the negative score is about 5(up to 10) and positive score is about 5(up to 10); if the text say, although bitcoin has high volatility, I like it, it should be higher positive score than negative score; these examples are make you understand how marking. A forementioned content is sample cryptocurrency sentiment analysis, the actual senario should be more complicated. For example the first text said: 1384888505 I’m one of the Senators attending today's U.S. Senate Banking Committee hearing related to bitcoin. What would you like me to know? The timestamp of this post is 1384888505, and then the content is related to the hearing, indicating that at that time, Bitcoin may have attracted a lot of attention, but at that time, people do not know whether this thing is good or not, and may have to hold a hearing. You need to consider the information at this level, and then assign positive and negative scores separately.
The following is the cryptocurrency scope top posts' text of reddit with its timestamp, you should analyze follow the aforementioned requirements and only give me the final score like "positivate score: 2, negative score: 3":

\section{Time Series Forecasting Model Descriptions}
\label{app:lstm and gru}

\ \ \ \textbf{ARIMA} \cite{shumway2017arima}: The AutoRegressive Integrated Moving Average (ARIMA) model is a classical statistical method used for analyzing and forecasting time series data. It combines autoregressive and moving average components and is effective for stationary time series.

\textbf{Linear Regression} \cite{zou2003correlation}: Linear Regression is a fundamental machine learning model that establishes a linear relationship between the input features and the target variable. It serves as a baseline for time series forecasting tasks.

\textbf{RNN} \cite{sherstinsky2020fundamentals}: Recurrent Neural Networks (RNNs) are a type of neural network designed for sequential data, where the hidden state captures temporal dependencies. RNNs are effective in modeling time series data but may suffer from vanishing gradient issues.

\textbf{MICN} \cite{wang2023micn}: The Multi-Input Convolutional Network (MICN) utilizes convolutional layers to capture local temporal patterns in time series data. It is particularly effective for short-term forecasting.

\textbf{TimesNet} \cite{wu2022timesnet}: TimesNet employs a hierarchical structure to model time series data, focusing on multi-resolution patterns and enhancing long-term prediction accuracy.

\textbf{TCN} \cite{hewage2020temporal}: Temporal Convolutional Networks (TCNs) are an alternative to RNNs for sequence modeling, offering parallelism and the ability to capture long-range dependencies through dilated convolutions.

\textbf{TimeMixer} \cite{wang2024timemixer}: TimeMixer integrates mixing operations across temporal dimensions to enhance feature interactions, making it effective for complex time series with diverse patterns.

\textbf{NLinear} \cite{zeng2023transformers}: NLinear simplifies the forecasting process by focusing on linear relationships in time series data, providing a lightweight yet effective solution.

\textbf{DLinear} \cite{qiu2024tfb}: DLinear extends NLinear by incorporating additional nonlinear components, improving its adaptability to complex data structures.

\textbf{PatchTST} \cite{nie2022time}: PatchTST employs patch-based representations and Transformer mechanisms to capture both local and global temporal features, enhancing time series forecasting.

\textbf{FITS} \cite{xu2023fits}: FITS (Forecasting via Interpretable Time Series) leverages interpretable Transformer structures to provide accurate and explainable forecasts.

\textbf{Crossformer} \cite{zhang2023crossformer}: Crossformer enhances temporal modeling by introducing cross-dimensional attention, which captures intricate dependencies within time series data.

\textbf{FEDformer} \cite{zhou2022fedformer}: FEDformer employs a frequency-enhanced approach to Transformer-based forecasting, improving accuracy in periodic and non-stationary time series.

\textbf{iTransformer} \cite{liu2023itransformer}: iTransformer introduces interactive attention mechanisms to enhance the model's ability to capture temporal relationships across diverse time scales.

\textbf{Non-stationary Transformer} \cite{liu2022non}: This model adapts the Transformer architecture to handle non-stationary time series, improving its robustness and generalization capabilities.

\textbf{Informer} \cite{zhou2021informer}: Informer optimizes the Transformer model for long-sequence forecasting by introducing sparse self-attention and enhanced positional encoding.

\textbf{Triformer} \cite{cirstea2022triformer}: Triformer is designed for triadic time series data, leveraging unique attention mechanisms to capture interactions among multiple sequences.

\textbf{FiLM} \cite{zhou2022film}: Feature-wise Linear Modulation (FiLM) is a model-agnostic method that dynamically adjusts feature representations to enhance model adaptability and performance.

\textbf{TimeGPT} \cite{garza2023timegpt}: TimeGPT is a large language model specifically pre-trained for time series data, leveraging its vast contextual understanding to deliver state-of-the-art forecasting performance.

\section{Price Prediction Experiment Setup}

The objectives of this experiment are, (1) comparing the performance of time series models in predicting the prices of single and multiple tokens, (2) testing the impact of additional features including global indices and sentiment, (3) evaluating the predictive performance of the models based on the token listing time. We divide the price prediction task into two parts:

\paragraph{Univariate forecasting data.}All models are applied to predict the prices of three tokens (LUNC, BTC, and ETC). For each model, we define three input vectors for evaluation:

\begin{itemize}
\item price\_{lunc}$^{\text{lunc}}_{j,t} = (p^{\text{lunc}}_{j,t})$
\item price\_{btc}$^{\text{btc}}_{j,t} = (p^{\text{btc}}_{j,t})$
\item price\_{eth}$^{\text{eth}}_{j,t} = (p^{\text{eth}}_{j,t})$
\end{itemize}

\paragraph{Multivariate forecasting data.}We selected 211 tokens with sufficiently long time series (e.g., Obyte, Blox, Augur, AppCoins) for Multivariate forecasting. These tokens have at least six years of price records. For each model, we define four input vectors for evaluation:

\begin{itemize}
\item price$_{j,t} = p_{j,t}$
\item price\_global$_{j,t} = (p_{j,t}, g_{j,t})$
\item price\_text$_{j,t} = (p_{j,t}, s^{\text{pos}}_t, s^{\text{neg}}_t)$
\item price\_all$_{j,t} = (p_{j,t}, g_{j,t}, s^{\text{pos}}_t, s^{\text{neg}}_t)$
\end{itemize}

\paragraph{Parameter Settings}

For univariate forecasting, we adopt a fixed forecasting strategy to maintain consistency with the M4 competition settings \cite{liu2022non, qiu2024tfb}. The forecasting horizon is set to 14, and the look-back window is set to 1.25 times the forecasting horizon. For multivariate forecasting, we use the rolling forecasting strategy with a forecasting horizon of 60 \cite{qiu2024tfb}.

For each method, we adhere to the hyper-parameters specified in their original papers. Additionally, we perform hyper-parameter searches across multiple sets, with a maximum of 8 sets tested. The optimal result is selected to ensure a comprehensive and unbiased evaluation of each method's performance.

All experiments are conducted using PyTorch \cite{paszke2019pytorch} in Python 3.10 and executed on an NVIDIA RTX 4090 GPU. The training process employs the L2 loss function and the ADAM optimizer \cite{diederik2014adam}. The initial batch size is set to 32, which can be halved (minimum of 8) in case of an Out-Of-Memory (OOM) error. The "Drop Last" operation is not used during testing. To ensure reproducibility, datasets and code are available at: \url{https://anonymous.4open.science/r/ICSWM-36BF/README.md}

\paragraph{Performance Metrics} 

For univariate forecasting, we use Normalized Mean Squared Error (MSE) and Normalized Mean Absolute Error (MAE) as performance metrics, which evaluate the error level and extreme error behavior of predictions. For multivariate forecasting, we use Normalized Weighted Absolute Percentage Error (WAPE) and Mean Symmetric Mean Absolute Percentage Error (MSMAPE) as performance metrics, which are suitable for reflecting overall relative error and the balance of positive and negative errors. Additionally, normalization eliminates the impact of scale differences across datasets, ensuring comparability and fairness in the evaluation of different tasks and models.

\section{LLM Knowledge Base Marked Sentiment Data Algorithm}
\label{algorithm}

Algorithm \ref{alg:textual_data_processing}, named LLM Knowledge Base Marked Sentiment Data, processes sentiment data from a CSV file to generate a DataFrame with updated values for scores, timestamps, number of comments, and sentiment attributes (positive and negative). The algorithm begins by reading and cleaning the data, then it processes each row by applying a decay factor \(k = 0.5\) to reduce the scores, comments, and sentiment values for certain time intervals. It also handles missing timestamps by filling in empty entries. The final output is a clean, processed DataFrame that includes adjusted sentiment information over a specified period.

\begin{algorithm}[H]
\caption{LLM Knowledge Base Marked Sentiment Data}
\label{alg:textual_data_processing}
\begin{algorithmic}
    \State \textbf{Input:} Path to the CSV file, decay factor $k = 0.5$
    \State \textbf{Output:} Processed DataFrame with updated score, timestamp, number of comments, positive and negative attributes

    \State Read the CSV file and clean the data
    \State Initialize an empty list $processed\_data$
    
    \For{each row in data frame}
        \State Extract $score$, $timestamp$, $number\_of\_comment$, $positive$, $negative$ from the row

        \For{$j \gets 0$ to $2$}
            \State Set $new\_timestamp = timestamp + j \cdot 1$ day
            \State Append $[score, new\_timestamp] $ to $processed\_data$
            \State Append $[number\_of\_comment] $ to $processed\_data$
            \State Append $[positive, negative]$to $processed\_data$
        \EndFor

        \For{$j \gets 3$ to $6$}
            \State $score \gets score \cdot k$
            \State $number\_of\_comment \gets number\_of\_comment \cdot k$
            \State $positive \gets positive \cdot k$
            \State $negative \gets negative \cdot k$
            \State Set $new\_timestamp = timestamp + j \cdot 1$ day
            \State Append $[score, new\_timestamp] $ to $processed\_data$
            \State Append $[number\_of\_comment]  $to $processed\_data$
            \State Append $[positive, negative]$to $processed\_data$
        \EndFor
        
        \If{next post timestamp $> timestamp + 7$ days}
            \While{last post day $<$ next post timestamp}
                \State Append $[0, last\_post\_day, 0, 0, 0]$ to $processed\_data$
                \State Increment last\_post\_day by 1 day
            \EndWhile
        \EndIf
    \EndFor

    \State Check for invalid timestamps and clean processed data
    \State Return $processed\_data$ as a DataFrame
\end{algorithmic}
\end{algorithm}

Algorithm~\ref{alg:common_sentiment_textual}, named Common LLM Marked Sentiment Data Processing, processes sentiment data from a CSV file to generate a DataFrame with updated attributes such as scores, timestamps, comment counts, and overall sentiment. The procedure starts by cleaning the data (converting timestamps, ensuring numeric columns, and removing missing values). It then processes each row, applying a decay factor \(k\) to adjust the scores and comments over time. The algorithm also fills in missing timestamps when necessary. Finally, it aggregates the data by summing scores and comments and averaging sentiment before returning the cleaned and sorted DataFrame.

\begin{algorithm}[H]
\caption{Common LLM Marked Sentiment Data Processing}
\label{alg:common_sentiment_textual}
\begin{algorithmic}
    \State \textbf{Input:} Path to CSV, decay factor $k$
    \State \textbf{Output:} Processed DataFrame with updated score, timestamp, number of comments, and sentiment

    \State \textbf{Step 1: Data Cleaning} 
    \State Convert `timestamp` to datetime, ensure columns are numeric, remove NaNs
    
    \State \textbf{Step 2: Data Processing} 
    \For{each row in the data}
        \State Extract features: $score$, $timestamp$, $number\_of\_comment$, $sentiment$
        \For{$j = 0$ to $2$}
            \State Update and append $score$ and $new\_timestamp$ to $processed\_data$
        \EndFor
        \For{$j = 3$ to $6$}
            \State Apply decay to $score$ and $number\_of\_comment$, append to $processed\_data$
        \EndFor
        \If{next post timestamp exceeds threshold}
            \State Append filler rows for missing days
        \EndIf
    \EndFor
    \State \textbf{Step 3: Aggregation and Averaging} 
    \State Group by `timestamp', sum `score' and `number\_of\_comment', average `sentiment' 
    \State \textbf{Step 4: Final Clean-up} 
    \State Sort by `timestamp', remove invalid entries, return $processed\_data$
\end{algorithmic}
\end{algorithm}

\

\end{document}